\documentclass[letterpaper,useAMS,usenatbib]{mn2e}
\usepackage{epsfig}

\newcommand{\ud}{{\rmn{d}}}
\newcommand{\vnhat}{\hat{\bmath{n}}}

\newcommand{\vb}{{\bmath{b}}}
\newcommand{\vh}{{\bmath{h}}}
\newcommand{\vE}{{\bmath{E}}}
\newcommand{\vl}{{\bmath{l}}}

\newcommand{\vy}{{\bmath{y}}}

\newcommand{\vx}{{\bmath{x}}}

\newcommand{\vs}{{\bmath{s}}}
\newcommand{\vp}{{\bmath{p}}}
\newcommand{\vnabla}{\bmath{\nabla}}
\newcommand{\vEA}{\bmath{E}_A}
\newcommand{\vEB}{\bmath{E}_B}
\newcommand{\vbA}{\bmath{b}_A}
\newcommand{\vbB}{\bmath{b}_B}
\newcommand{\vbd}{\bmath{b}_d}
\newcommand{\mM}{{\mathbfss{M}}}
\newcommand{\mI}{{\mathbfss{I}}}
\newcommand{\mR}{{\mathbfss{R}}}
\newcommand{\mJ}{{\mathbfss{J}}}
\newcommand{\mF}{{\mathbfss{F}}}
\newcommand{\mLam}{\bmath{\Lambda}}

\newcommand{\half}{{\textstyle{\frac{1}{2}}}}
\newcommand{\quart}{{\textstyle{\frac{1}{4}}}}

\newcommand{\Muller}{M\"{u}ller\ }
\newcommand{\dualvEA}{{}^\star\!\bmath{E}_A}
\newcommand{\dualvEB}{{}^\star\!\bmath{E}_B}
\newcommand{\EAco}{E_{A\,\mathrm{co}}}
\newcommand{\EAcross}{E_{A\,\mathrm{cross}}}
\newcommand{\EBco}{E_{B\,\mathrm{co}}}
\newcommand{\EBcross}{E_{B\,\mathrm{cross}}}
\newcommand{\CT}{C^T_b}
\newcommand{\CE}{C^E_b}
\newcommand{\CB}{C^B_b}
\newcommand{\CBlens}{C^B_{b,\mathrm{lens}}}

\newcommand{\CTsq}{(C^T_b)^2}
\newcommand{\CEsq}{(C^E_b)^2}
\newcommand{\CBsq}{(C^B_b)^2}
\newcommand{\CBCE}{C^B_b C^E_b}
\newcommand{\CBCT}{C^B_b C^T_b}
\newcommand{\CBCET}{C^B_b C^{TE}_b}
\newcommand{\hatC}{\hat{C}}
\newcommand{\CTE}{C^{TE}_b}
\newcommand{\CBobs}{\hat{C}^B_{b,\mathrm{obs}}}
\newcommand{\var}{\mathrm{var}}

\newcommand{\lssq}{\sigma^2 l^2}
\newcommand{\lsfth}{\sigma^4 l^4}
\newcommand{\lssix}{\sigma^6 l^6}
\newcommand{\lseth}{\sigma^8 l^8}
\newcommand{\nuar}{\nu_{a,R}}
\newcommand{\nudr}{\nu_{d,R}}
\newcommand{\nuai}{\nu_{a,I}}
\newcommand{\nudi}{\nu_{d,I}}
\newcommand{\plusp}{{}_1 p}
\newcommand{\minusp}{{}_{-1} p}
\newcommand{\plusbd}{{}_1 b_d}
\newcommand{\minusbd}{{}_{-1} b_d}
\newcommand{\cle}{\mathcal{E}}

\title[Systematic errors in CMB
polarization measurements]
{Systematic errors in cosmic microwave background polarization measurements}
\author[Daniel O'Dea, Anthony Challinor and Bradley R. Johnson]
{Daniel O'Dea,$^1$\thanks{E-mail: dto22@mrao.cam.ac.uk}
Anthony Challinor$^{1,2,3}$\thanks{E-mail: a.d.challinor@ast.cam.ac.uk}
and Bradley R. Johnson$^4$\thanks{E-mail: bjohnson@physics.ox.ac.uk}
\\
$^1$Astrophysics Group, Cavendish Laboratory, J.J.~Thomson Avenue,
Cambridge CB3 0HE, U.K.
\\
$^2$Institute of Astronomy, Madingley Road, Cambridge, CB3 0HA, U.K.
\\
$^3$DAMTP, Centre for Mathematical Sciences, Wilberforce Road,
Cambridge, CB3 0WA, U.K.
\\
$^4$Oxford Astrophysics, Department of Physics, Denys Wilkinson Building,
Keble Road, Oxford, OX1 3RH, U.K.
}

\date{Accepted 2007 January 28. Received 2007 January 24; in original
  form 2006 October 12}

\pagerange{\pageref{firstpage}--\pageref{lastpage}}
\pubyear{2007}

\label{firstpage}

\begin{document}

\maketitle

\begin{abstract}
We investigate the impact of instrumental systematic errors on the
potential of cosmic microwave background polarization experiments
targeting primordial $B$-modes. To do so, we introduce spin-weighted
M\"{u}ller matrix-valued fields describing the linear response of
the imperfect optical system and receiver, and give a careful
discussion of the behaviour of the induced systematic effects under
rotation of the instrument. We give the correspondence between the
matrix components and known optical and receiver imperfections, and
compare the likely performance of pseudo-correlation receivers and
those that modulate the polarization with a half-wave plate. The
latter is shown to have the significant advantage of not coupling
the total intensity into polarization for perfect optics, but
potential effects like optical distortions that may be introduced
by the quasi-optical wave plate warrant further investigation.
A fast method for tolerancing time-invariant
systematic effects is presented, which propagates errors through to
power spectra and cosmological parameters. The method extends
previous studies to an arbitrary scan strategy, and eliminates the
need for time-consuming Monte-Carlo simulations in the early phases
of instrument and survey design. We illustrate the method with both
simple parametrized forms for the systematics and with beams based
on physical-optics simulations. Example results are given in the
context of next-generation experiments targeting tensor-to-scalar
ratios $r\sim 0.01$. \vspace{\baselineskip}
\end{abstract}

\begin{keywords}
cosmic microwave background -- methods: analytical -- methods: numerical.
\end{keywords}

\section{Introduction}
\label{sec:introduction}

There is currently a great deal of interest and activity in the
field of cosmic microwave background
(CMB) polarimetry. Several recent
experiments~\citep{DASI,CBIpol,CAPMAP,boomerangEE,WMAP3pol, 2006astro.ph.11392W} have
reported detections of CMB polarization at a level consistent
with predictions in simple cold dark matter (plus $\Lambda$) models fit
to the temperature anisotropy data. In standard adiabatic models,
polarization measurements promise a further tightening of constraints
inferred from the temperature anisotropies, and the breaking of
important degeneracies~\citep*{ZSS}. In particular, with the recent
release of the three-year \emph{Wilkinson Microwave Anisotropy Probe}
(\emph{WMAP})
data, the optical depth to reionization has been constrained to
$\tau = 0.09 \pm 0.03$~\citep{WMAP3pe} with the large-angle
polarization signal~\citep{Zalri}. Polarization information is
particularly valuable in non-standard models, for example it can
significantly improve constraints in the presence of isocurvature
modes~\citep*{iso}, and be used to constrain parity-violating
physics
(e.g.~\citealt{1997PhRvD..56.7493S}; \citealt*{1999PhRvL..83.1506L}; \citealt{2006PhRvL..96v1302F}).
Secondary effects, most notably weak gravitational
lensing~\citep{1998PhRvD..58b3003Z,2006PhR...429....1L}, further encode
information on the low-redshift universe in CMB polarization.

Perhaps the most exciting aspect of CMB polarimetry is the window it
may open on primordial gravitational waves~(\citealt*{KKS}; \citealt{ZS97}).
The curl mode (or $B$-mode) of polarization is not produced
by linear density perturbations and so provides an observable
signature of gravitational waves at last scattering that is not limited
by cosmic variance from the dominant density perturbations. However,
the situation is clouded by several issues. First, the amplitude
of gravitational waves expected from inflation is unknown as the energy scale
at which inflation may have occurred is not (yet) determined by theory.
What is known is that any $B$-mode imprint will be very small: the
current limit on the amplitude of the gravitational wave power spectrum
(expressed as a fraction of that for the density perturbation) is
$r < 0.28$ at 95\%\ confidence~\citep{WMAP3pe}, from a combination
of the temperature anisotropies and galaxy-clustering data. This limit on
$r$ translates to an r.m.s.\ $B$-mode signal $< 200\,\mathrm{nK}$.
Furthermore,
gravitational lensing \emph{does} produce $B$-mode polarization at second
order in the density perturbations, with a spectrum that is almost white
for multipoles $l \la 300$. The amplitude $\sqrt{C^B_{l, \mathrm{lens}}}
\approx 5\, \mu$K-arcmin means the lens-induced $B$-modes dominate the
primordial ones for $r \la 0.01$ except on very large scales where
the latter is enhanced by reionization. For noise levels much better than
$5\, \mu$K-arcmin, it will be worthwhile to try and clean out the
large-angle $B$-modes of lensing by making use of
non-Gaussianity (e.g.~\citealt{2002ApJ...574..566H}). Finally,
polarized emission from Galactic foregrounds will likely overwhelm the
primordial $B$-mode signal over a large fraction of the
sky~\citep{WMAP3pol}, and
our ability to constrain gravitational waves will almost certainly
be limited by the accuracy at which Galactic foregrounds can
be subtracted.

The intrinsic weakness of the polarization signal presents a major
experimental challenge. The gradient (or $E$-mode) polarization is now
determined to be at least an order of magnitude smaller than
the temperature anisotropies for $l \la 1000$, and the $B$-modes (including
lensing) are indirectly constrained to be at least an order of
magnitude smaller still. As well as the raw
sensitivity requirements that this implies, any systematic
errors present in $B$-mode instruments will have to be controlled to an
unprecedented level of accuracy in order to ensure the signal is not
fatally contaminated. With this motivation, in this paper we
develop a general framework to describe and assess the impact of
instrumental systematic errors on experiments targeted at CMB polarization.

There is a growing literature discussing the impact of systematic
effects in CMB polarimetry; for discussions of the real-world issues
encountered in recent surveys,
see~\citet{2005ApJS..159....1B,2005astro.ph..7509M,WMAP3pol,2006astro.ph..6606J}.
There are typically many sources of potential systematic error. Here
we shall be concerned with the broad class of errors that produce a
time-independent residual signal for a given instrument pointing.
These include imperfections in the receiver and optics, but exclude
important effects such as low-frequency noise from thermal
fluctuations, readout electronics or atmospheric fluctuations, and
pointing jitter. Low-frequency noise is best dealt with by a
combination of signal modulation (active or by scanning), a well
cross-linked survey, and removal during the map-making stage. Our
aim is twofold: to give a careful analysis of the transformation
properties of various systematic effects under rotation of the
instrument -- a useful strategy for mitigating some systematics; and
to provide a fast, semi-analytic method for tolerancing systematic
effects that is flexible enough to deal with arbitrary scan
strategies and removes the need for time-consuming simulations
during the early stages of instrument and survey design. Our
approach is similar to that of~\citet*{Hu03}, but we extend their
analysis in several ways. We describe the receiver systematics with
M\"{u}ller matrices and the optics with M\"{u}ller matrix-valued
fields which allows us to deal with arbitrary beam effects. To
simplify the discussion of the rotation properties of systematic
effects, we introduce spin-weighted M\"{u}ller matrices and perform
a further decomposition of the beam matrix-valued fields into
irreducible components. The latter singles out those features of the
beams that cannot be overcome with instrument rotation, and easily
reproduces the results of~\citet{2004A&A...420..437C} for
axisymmetric systems. In addition, our analysis of the scientific
impact of systematic effects works with an arbitrary scan strategy,
and propagates the effects through to biases in cosmological
parameters and the increase in their random errors. Carrying biases
through to parameters is potentially important: systematic errors
often must be suppressed further than the cosmic-variance limit in
the CMB power spectra if they are to have negligible impact on the
science~\citep{1999MNRAS.304...75E}. Also, properly modelling the
interaction between systematic effects and the scan strategy is
important as its effects cannot easily be predicted, or constrained,
from simpler scans (such as a raster scan with no implied beam
rotation). For example, a non-trivial scan can cause leakage from
the dominant $E$-modes into $B$-modes that is not present for a
raster scan, as a result of the non-local nature of the $E$- and
$B$-mode decomposition.

This paper is organized as follows. Section~\ref{sec:notation}
introduces our notation and some important assumptions. Propagation
through the receiver is described in
Section~\ref{sec:mullermatrices} in terms of spin-weighted \Muller
matrices, while the optics are described in terms of matrix-valued
fields in Section~\ref{sec:mullerfields}. We give illustrations for
common receiver types and optical imperfections, and compare their
relative merits. In Section~\ref{sec:powerspectra} we present our
method for propagating errors to biases in power spectra and the
enhancement of their variances, and follow these through to
cosmological parameters (specifically $r$) in
Section~\ref{sec:biases}. Finally, in~Section~\ref{sec:applications}
we apply our methods to set tolerances on parametrized errors and
compare these for different scan strategies. We also examine the
impact of realistic, simulated beam profiles. Two appendices give
further properties of  the complex \Muller matrices and of the beam
expansion in irreducible components introduced in
Section~\ref{sec:mullerfields}.

\section{Notation}
\label{sec:notation}

For surveys covering a small fraction of the sky we can work in the flat-sky
limit where the CMB fields are considered on the tangent plane to the
celestial sphere. We adopt Cartesian coordinates on this plane such
that $x$ points north--south and $y$ points west--east; the right-handed
$z$-direction is then along the line of sight.
We define Stokes parameters using the $x$ and $-y$ directions for which
the radiation propagation direction completes a right-handed triad. Our
polarization conventions are then consistent with the IAU standards.
If the electric field of the incident radiation at frequency
$\omega$ is $\Re (\vE e^{-i\omega t})$,
with these conventions we have
\begin{equation}
\langle E_i E_j^* \rangle = \frac{1}{2}\left( \begin{array}{cc}
T+Q & -U - iV \\
-U + iV & T - Q
\end{array} \right) .
\label{eq:not1a}
\end{equation}
We shall only consider quasi-monochromatic systems here but our discussion
will still hold for broad-band systems if the non-ideal instrument
responds in the same way to all in-band radiation. Clearly this
ignores an important class of systematic effects, for example
bandpass mismatch~\citep{2006astro.ph..3452J}.

The complex polarization $P(\vx) \equiv (Q+iU)(\vx)$ is spin $-2$ in
the sense that under the transformation $\hat{\vx}+i\hat{\vy} \mapsto
(\hat{\vx}+i\hat{\vy}) e^{i\psi}$, which rotates the $x$-axis by $\psi$
towards the \emph{negative} $y$-axis, $P\mapsto P e^{-2i\psi}$. Decomposing
$P$ into its electric (E) and magnetic (B) parts, we have, in Fourier space,
\begin{equation}
(Q\pm i U)(\vx) = - \int \frac{\ud^2\vl}{2\pi}[E(\vl)\mp iB(\vl)]
e^{\mp 2i\phi_\vl} e^{i\vl\cdot \vx},
\label{eq:not1}
\end{equation}
where $\vl = l(\cos\phi_\vl,\sin\phi_\vl)$. For the cosmological
examples we give in this paper we only consider the CMB fields
on the sky. For observations in the quiet Galactic regions that will
be the targets for future ground-based $B$-mode experiments,
the CMB temperature fluctuations and $E$-mode polarization should
dominate Galactic emission. Ignoring the latter should therefore
be harmless for those most troubling systematic effects that couple
temperature and $E$-mode polarization into $B$.

\section{Receiver M\"{u}ller matrices}
\label{sec:mullermatrices}

The \Muller matrices describe the propagation of the Stokes parameters
through the receiver element of a given observing system.
The optical coupling to the fields on the sky requires a
description in terms of M\"{u}ller matrix-valued
fields which we describe in Section~\ref{sec:mullerfields}.

Gathering
the Stokes parameters in a Stokes vector $\vs \equiv (T,Q,U,V)^T$,
we have for the observed Stokes vector
\begin{eqnarray}
\vs_{\mathrm{obs}} &=&
\left(\begin{array}{cccc} M_{TT} & M_{TQ} & M_{TU} & M_{TV} \\
              M_{QT} & M_{QQ} & M_{QU} & M_{QV} \\
                  M_{UT} & M_{UQ} & M_{UU} & M_{UV} \\
              M_{VT} & M_{VQ} & M_{VU} & M_{VV}  \end{array}
\right) \vs.
\label{eq:muller1}
\end{eqnarray}
We adopt the convention here that \Muller matrices are always expressed
in the instrument basis. This coincides with the
Cartesian sky basis ($x$ and $-y$) when the
instrument is in its \emph{fiducial orientation}. It is convenient to
work with the complex \Muller matrix whose elements have definite spin, i.e.\
\begin{eqnarray}
\vp_{\mathrm{obs}} &=&
\left(\begin{array}{cccc} M_{TT} & M_{TP} & M_{TP^*} & M_{TV} \\
              M_{PT} & M_{PP} & M_{PP^*} & M_{PV} \\
                  M_{P^* T} & M_{P^* P} & M_{P^* P^*} & M_{P^* V} \\
              M_{VT} & M_{VP} & M_{VP^*} & M_{VV} \end{array}
\right)
\vp.
\label{eq:muller2}
\end{eqnarray}
Here, the complex Stokes vector is $\vp=(T,P,P^*,V)^T$. We denote
the matrix on the right of equation~(\ref{eq:muller2}) by $\mM$. Its
components are related to those in equation~(\ref{eq:muller1}) as follows:
for the total intensity
\begin{equation}
M_{TP} = \half (M_{TQ} - i M_{TU}), \quad
M_{TP^*} = \half (M_{TQ} + i M_{TU}),
\label{eq:muller3}
\end{equation}
with analogous results for $M_{VP}$ and $M_{VP^*}$;
and for the polarization
\begin{equation}
M_{PT} = M_{QT} + i M_{UT} , \quad  M_{PV} = M_{QV} + i M_{UV},
\label{eq:muller4}
\end{equation}
and
\begin{eqnarray}
M_{PP} &=& \half (M_{QQ}+M_{UU}) + \half i (M_{UQ}-M_{QU}) , \nonumber \\
M_{PP^*} &=& \half (M_{QQ}-M_{UU}) + \half i (M_{UQ}+M_{QU}).
\label{eq:muller5}
\end{eqnarray}
The components for the spin-2 polarization $P^*$ are related to those
for $P$:
\begin{eqnarray}
M_{P^*T}=M_{PT}^* ,\quad && M_{P^*V}=M_{PV}^* , \nonumber \\
M_{P^*P}=M_{PP^*}^*, \quad &&  M_{P^*P^*}=M_{PP}^* .
\label{eq:muller6}
\end{eqnarray}

If we kept the instrument in a fixed orientation, but transformed to a rotated
basis in describing the polarization fields on the sky, the \Muller
matrix elements would transform like the complex conjugate of the field
appearing in the second index, e.g.\ $M_{PP} \mapsto M_{PP} e^{2i\psi}$
so that $M_{PP}P$ remained constant. If we further rotated the observed
polarization from the instrument basis to the rotated sky basis, we would
pick up an additional factor $e^{is\psi}$ where $s$ is the spin of the
field associated with the first index. More relevant for our purposes are
the transformation properties of the \Muller matrix under (active) rotations
of the instrument. Let us rotate the instrument by $\psi$ taking $x$ towards
the negative $y$-axis and simultaneously back-rotate the observed polarization
so we are describing the measured polarization in the original sky basis.
In its basis, the instrument sees incoming radiation with complex Stokes
vector $\mLam(\psi)\vp$, where $\mLam(\psi)\equiv \mathrm{diag}(1,e^{-2i\psi},
e^{2i\psi},1)$, so the observed polarization on the sky basis is
$\vp_{\mathrm{obs}}(\psi)=\mLam^\dagger(\psi)\mM\mLam(\psi)\vp$.
Note that only the diagonal elements of $\mM$ are invariant under
$\mM \mapsto \mLam^\dagger(\psi)\mM\mLam(\psi)$.

In the case of an ideal instrument, $\mM$ is equal to the
identity matrix and any systematic errors that affect the Stokes
parameters will lead to small perturbations from this. It is convenient
to introduce a more concise notation for those elements describing
coupling to $P$ and $P^\ast$~\citep{Hu03}: $M_{PT} = \gamma_1 + i \gamma_2$,
$M_{PP} = 1+ a + 2i\omega$, $M_{PP^\ast} = f_1 + i f_2$ and
$M_{PV} = w_1 + i w_2$, where all parameters on the right-hand sides
are real.
For the perturbation to the linear polarization,
we then have
\begin{eqnarray}
\delta(Q\pm iU) &=& (a\pm 2i\omega)(Q\pm iU)+(f_1\pm if_2)(Q\mp iU)
\nonumber \\
&&\mbox{} + (\gamma_1 \pm i\gamma_2)T + (w_1 \pm i w_2)V.
\label{eq:muller7}
\end{eqnarray}
These parameters are defined in the fiducial basis, and describe a
miscalibration  of the polarization amplitude, $a$, a rotation of
the polarization orientation, $\omega$, transformations between the
two polarization spin states, $f_1$ and $f_2$, leakage from total
intensity to $Q$ with $\gamma_1$ and to $U$ with $\gamma_2$, and
leakage from circular polarization with $w_1$ and $w_2$. The circular
leakage terms were
not considered in~\citet{Hu03}. Under a rotation of the instrument
as described above, we find
\begin{eqnarray}
\delta(Q\pm iU) &=&
(a\pm 2i\omega)(Q\pm iU)                              \nonumber \\
&& \mbox{} + (f_1\pm if_2)(Q\mp iU) e^{\pm 4i\psi}    \nonumber \\
&& \mbox{} + (\gamma_1 \pm i\gamma_2)T e^{\pm 2i\psi} \nonumber \\
&& \mbox{} + (w_1 \pm i w_2)V e^{\pm 2i\psi}.
\label{eq:muller9}
\end{eqnarray}
This suggests that the $f$, $\gamma$ and $w$ errors can be controlled
with instrument rotation as these terms have different spin properties
to the fields they are perturbing.
For example, if we average observations
of a point over all possible orientations these terms
disappear. We look at controlling systematics with beam rotation
in more detail in Section~\ref{sec:calibration}.

\subsection{Receiver errors}
\label{sec:receiver}

\begin{figure}
\begin{center}
  \includegraphics[width=84mm]{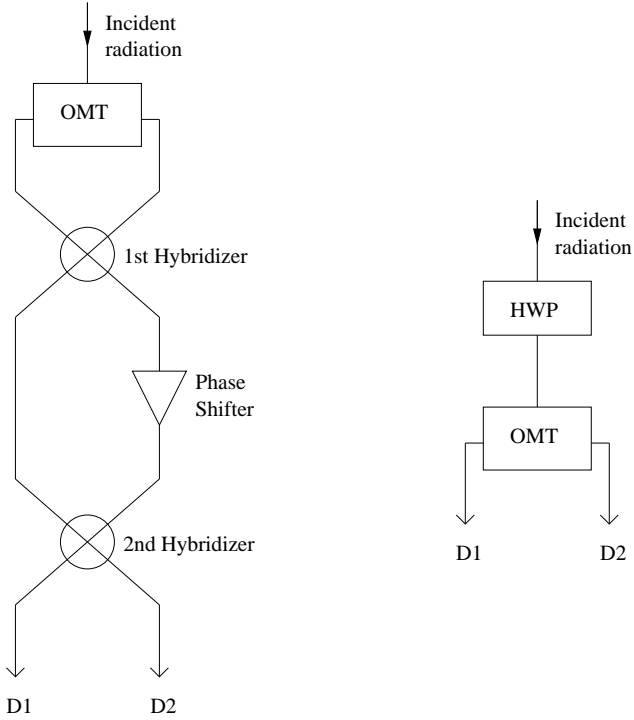}
  \caption[Receiver block diagrams]{Block diagrams for a
    pseudo-correlation receiver (left) and a rotating half-wave-plate
    receiver (right). For the pseudo-correlation receiver, the incident
    radiation is split into two orthogonal components by the ortho-mode
    transducer (OMT), and then
    propagates through a $90^\circ$ hybridizer. A time dependent phase
    shift is introduced along one arm, and the radiation passes
    through a second $90^\circ$ hybridizer before being detected. For the
    rotating half-wave-plate receiver, the incident radiation simply
    passes through a rotating half-wave plate before being split by the
    OMT and detected.}
  \label{fig:receiver}
\end{center}
\end{figure}

In this section we give examples of the polarization M\"{u}ller matrix
elements for two common polarimeters that have particular relevance to CMB
polarimetry: the pseudo-correlation receiver and the rotating
half-wave-plate receiver. Block diagrams for these receivers are
shown in Fig. \ref{fig:receiver}.

Quite generally, the propagation of radiation through a receiver can be
described by a Jones matrix, $\mJ$, such that the electric field after
passing through the receiver, $\vE_{\mathrm{rec}}$, is
\begin{equation}
\vE_{\mathrm{rec}}=\mJ \vE,
\label{eq:rec1}
\end{equation}
where $\vE$ is the incident electric field. In this section only,
the elements of $\vE$ are
the complex amplitudes of the two linear polarizations, $A$ and $B$, which,
for an ideal optical system, couple in the far-field to the $x$ and $-y$
components of the electric field of the incident radiation; see
Section~\ref{sec:mullerfields}.
For a receiver with several components, the Jones matrix of the
receiver is the product of the matrices for each component provided that
reflections can be ignored. The \Muller matrix for the receiver
can be found from the relations $\vs_{\mathrm{obs}} = \mM \vs$ and
\begin{eqnarray}
\left( \begin{array}{cc} T+Q & U + i V \\
             U-iV & T-Q \end{array} \right)_{\mathrm{obs}}
&=& \mJ \left( \begin{array}{cc} T+Q & U + i V \\
             U-iV & T-Q \end{array} \right) \mJ^\dagger. \nonumber\\
&&\label{eq:rec1a}
\end{eqnarray}

For the pseudo-correlation receiver, the ideal Jones matrix is
\begin{eqnarray}
\mJ_{\mathrm{pc}} &=&
\mJ_{\mathrm{hybrid,2}}  \mJ_{\mathrm{phase}}
\mJ_{\mathrm{hybrid,1}}  \mJ_{\mathrm{omt}} \nonumber \\
&=&  \frac{1}{\sqrt{2}}
\left( \begin {array}{cc} 1&i\\ i&1\end {array} \right)
\left( \begin {array}{cc} 1&0\\ 0&e^{i\varphi t}\end {array} \right)
\nonumber \\
&&\mbox{} \times
\frac{1}{\sqrt{2}}
\left( \begin {array}{cc} 1&i\\ i&1\end {array} \right)
\left( \begin {array}{cc} 1&0\\ 0&1\end {array} \right) \nonumber\\
&=& \frac{1}{2}
\left( \begin {array}{cc} 1-e^{i\varphi t}&i(1 + e^{i\varphi t})\\
 i(1+e^{i\varphi t})&-1+e^{i\varphi t}\end {array} \right),
\label{eq:rec2}
\end{eqnarray}
where $\varphi t$ is the time-dependent phase shift, assumed to be
continuous here.
After passing through
the receiver, detectors measure the power in the two components,
$E_{\mathrm{rec},A}$ and $E_{\mathrm{rec},B}$ and, in the ideal,
noiseless case, their outputs for a particular pointing are
\begin{eqnarray}
D_1 &\equiv& \langle |E_{\mathrm{rec},A}|^2 \rangle =
\half(T - Q \cos{\varphi t} - U \sin{\varphi t} ) \nonumber\\
D_2 &\equiv& \langle |E_{\mathrm{rec},B}|^2 \rangle =
\half(T + Q \cos{\varphi t} + U \sin{\varphi t}).
\label{eq:rec3}
\end{eqnarray}
The total intensity and linear polarization Stokes parameters
can be recovered by taking
the sum and difference of the detector outputs and demodulating.
If we assume that the modulation frequency is much higher than the
maximum frequency in the incident signal (which is determined by the
scan speed and resolution), we can approximate the demodulation step by
\begin{eqnarray}
T_{\mathrm{obs}} &=& \frac{1}{\Delta} \int_0^\Delta
(D_1 + D_2)\,dt \nonumber\\
Q_{\mathrm{obs}} &=& \frac{2}{\Delta}  \int_0^\Delta
(D_2 - D_1)\cos{\varphi t}\,dt \nonumber\\
U_{\mathrm{obs}} &=& \frac{2}{\Delta}  \int_0^\Delta
(D_2 - D_1)\sin{\varphi t}\,dt ,
\label{eq:rec4}
\end{eqnarray}
where $\Delta$ is long compared to $1/\varphi$. We can now follow
systematic errors introduced in the Jones matrices through to the
observed Stokes parameters. We parametrize the systematic errors in
the various receiver components as follows:
\begin{eqnarray}
\mJ_{\rm{omt}} &=& \left( \begin {array}{cc} 1+g_1&\epsilon_1e^{i\theta_1}\\
  \epsilon_2e^{i\theta_2}&(1+g_2)e^{i\alpha}\end {array} \right) \nonumber\\
\mJ_{\rm{phase}} &=&  \left( \begin {array}{cc} 1&0\\
  0&e^{i(\varphi t + \delta \phi)}\end {array} \right) \nonumber\\
\mJ_{\rm{hybrid},j} &=& \frac{1}{\sqrt{2}} \left(
  \begin {array}{cc}(1+A_j)e^{ia_j}&i(1+B_j)e^{ib_j} \\
  i(1+C_j)e^{ic_j}&(1+D_j)e^{id_j}\end {array} \right),
\label{eq:rec5}
\end{eqnarray}
where $j=1$, $2$ labels the two hybridizers. Each parameter corresponds
to a potential, physical systematic error, for example, $g_1$ and $g_2$
represent gain errors in the two arms of the OMT.
Following the same process as in the ideal case, we find the effect of
these errors on the observed Stokes parameters. Assuming that
the systematic errors do not vary significantly over the time of the
observation, and expanding to linear order, the errors are related to the
parameters introduced in equation~(\ref{eq:muller7}) by
\begin{eqnarray}
a &=& \half(A_1+B_1+C_1+D_1+A_2+B_2+C_2+D_2)\nonumber\\
& & + g_1 + g_2 \nonumber \\
2\omega &=& \half(a_1+b_1-c_1-d_1+a_2+c_2-b_2-d_2) \nonumber\\
& & + \epsilon_2\cos{\theta_2}-\epsilon_1\cos{\theta_1}-\delta \phi\nonumber \\
\gamma_1 &=& g_1-g_2+\half(A_1+C_1-B_1-D_1) \nonumber\\
\gamma_2 &=&
\epsilon_1\cos{\theta_1}+\epsilon_2\cos{\theta_2}+\half (a_1+d_1-b_1-c_1)\nonumber \\
f_1 &=& 0 \nonumber\\
f_2 &=& 0 \nonumber\\
w_1 &=&  \half
(B_1+C_1-A_1-D_1)+\epsilon_1\sin{\theta_1}+\epsilon_2\sin{\theta_2}
\nonumber\\
w_2 &=&  \half (b_1 + d_1 - a_1 -c_1 + 2\alpha).
\label{eq:rec6}
\end{eqnarray}
As expected, we see that differential gain errors $g_1 - g_2$ lead
to instrumental $Q$ polarization, $\gamma_1$. `Spin-flip' errors,
coupling $P$ to $P^\ast$, are absent at first order but appear at
second order in the perturbation and higher. It should be noted that
the validity of the perturbative expansion depends in part on the
relative amplitudes of the polarization and total-intensity fields.
For example, we are implicitly assuming that any parameter that
contributes to $a$ at first order and to $\gamma_1$ at only second
order is sufficiently small to suppress the total-intensity leakage
caused to well below the level of the polarization leakage.

For the half-wave-plate receiver, with the plate rotating at an
angular velocity $\varphi$, the ideal Jones matrix is
\begin{eqnarray}
\mJ_{\mathrm{rhwp}} &=&
\mJ_{\mathrm{omt}}  \mJ^\mathrm{T}_{\mathrm{rot}}
\mJ_{\mathrm{hwp}}  \mJ_{\mathrm{rot}} \nonumber \\
&=&  \left( \begin {array}{cc} 1&0\\
0&1\end {array} \right)
\left( \begin {array}{cc} \cos{\varphi t}&-\sin{\varphi t}\\
\sin{\varphi t}&\cos{\varphi t}\end {array} \right) \nonumber\\
&&\mbox{} \times  \left( \begin {array}{cc} 1&0\\
0&-1\end {array} \right)
\left( \begin {array}{cc} \cos{\varphi t}&\sin{\varphi t}\\
-\sin{\varphi t}&\cos{\varphi t}\end {array} \right) \nonumber\\
&=& \left( \begin {array}{cc}\cos{2\varphi t} &\sin{2\varphi t}\\
\sin{2\varphi t}&-\cos{2\varphi t}\end {array} \right).
\label{eq:rec7}
\end{eqnarray}
This leads to similar
ideal detector outputs as the pseudo-correlator,
but with $Q$ and $U$ modulated at a frequency of $4\varphi$:
\begin{eqnarray}
D_1 &=& \half(T+Q \cos 4\varphi t + U \sin 4 \varphi t ) \nonumber \\
D_2 &=& \half(T-Q \cos 4\varphi t - U \sin 4 \varphi t ) .
\end{eqnarray}
Systematic errors in the OMT are parametrized as in
equation~(\ref{eq:rec5}), and for the other components,
\begin{eqnarray}
 \mJ_{\mathrm{hwp}} &=& \left( \begin {array}{cc} 1+h_1&\zeta_1e^{i\chi_1}\\
\zeta_2e^{i\chi_2}&-(1+h_2)e^{i\beta}\end {array} \right) \nonumber\\
\mJ_{\mathrm{rot}} &=& \left( \begin {array}{cc} \cos{(\varphi
    t+\delta\phi)}&\sin{(\varphi t+\delta\phi)}\\
-\sin{(\varphi t+\delta\phi)}&\cos{(\varphi t+\delta\phi)}\end {array} \right).
\label{eq:rec8}
\end{eqnarray}
Propagating these errors through to the observed Stokes parameters, we find
the only non-zero polarization couplings are
\begin{eqnarray}
a        &=& g_1+g_2+h_1+h_2 \nonumber \\
2\omega  &=& \epsilon_1\cos{\theta_1}-\epsilon_2\cos{\theta_2}-4\delta
\phi \nonumber \\
&&\mbox{} - \zeta_1\cos{\chi_1}-\zeta_2\cos{\chi_2} .
\label{eq:rec9}
\end{eqnarray}
The observation that $P$ couples only to $P$ actually holds exactly
for this receiver, and not just to first order.
By comparing equations (\ref{eq:rec6}) and (\ref{eq:rec9}) we can
begin to draw some useful conclusions as to the relative suitability
of these receivers for CMB polarimetry.
The half-wave-plate receiver has the potentially significant
advantage of having no total intensity leakage, given the assumptions made.
The large difference in the amplitude of the temperature and
polarization signals means that such leakage is potentially very
damaging, and hence any systematic errors that contribute to $\gamma_1$
and $\gamma_2$ will have very strict tolerance limits. It should be
noted that, in the presence of realistic optics, it is likely
that systematic errors in  the half-wave-plate receiver will
contribute to such leakage, but not at first order, as seen for the
pseudo-correlation receiver. Also, for a quasi-optical wave plate,
further investigation is still required into potential effects such as
optical distortions introduced by the plate \citep{2006astro.ph.11394J}.

\section{Beam M\"{U}ller fields}
\label{sec:mullerfields}

The \Muller matrices of the previous section describe propagation
of the Stokes parameters through the receiver. The propagation through
the entire instrument, including the optics, is described by matrix-valued
fields.
These couple the polarization fields on the sky to the observed
polarization that we assign to the nominal pointing direction. If we ignore
the effects of reflections from different stages of the receiver (and
the associated standing waves they set up), we can multiply the \Muller
matrices for the various stages and the optics to get the matrix describing
transfer through the entire system. In this section we consider the \Muller
matrix associated with the instrument optics (and horn, if present).

For a dual-polarization system, let the two orthogonal polarizations have
associated far-field radiation patterns $\vE_A(\vnhat)$ and $\vE_B(\vnhat)$.
It is convenient to define the fiducial orientation such that the nominal
directions on the sky defined by $\vE_A$ and $\vE_B$
(i.e.\ their co-polar parts) are along the $x$ and negative $y$-axis of the sky
coordinate system, which, recall, is the basis we are using to define
Stokes parameters. Of course, for a non-ideal system there will also be
cross-polar components of $\vE_A$ and $\vE_B$
perpendicular to the co-polar components. If we adopt the Ludwig-III
standard for the co- and 
cross-polar basis, the cross-polar direction for $A$ is along $-y$ and
for $B$ is along $-x$. We shall express the components of the radiation
patterns in terms of co- and cross-polar (complex) amplitudes.
The coupling of polarization $A$ to the incident radiation along a line of
sight $\vnhat$ is $\propto \vE_A(\vnhat) \cdot \vE(\vnhat)$ and similarly
for $B$.
The coherency of these signals defines Stokes parameters that are weighted
integrals of the Stokes parameters on the sky and from which we can extract
the beam \Muller fields:
\begin{eqnarray}
M_{TT} &=& \half (|\vE_A|^2 + |\vE_B|^2) \nonumber \\
M_{TQ} &=& \half (|\EAco|^2 - |\EAcross|^2 + |\EBcross|^2 - |\EBco|^2)
 \nonumber \\
M_{TU} &=& \half(\EAco \EAcross^* - \EBco \EBcross^*) + \mathrm{c.c.}
\nonumber \\
M_{TV} &=& \half i(\EAco\EAcross^* + \EBco\EBcross^*) + \mathrm{c.c.} \nonumber
\\
M_{QT} &=& \half (|\vE_A|^2 - |\vE_B|^2) \nonumber \\
M_{QQ} &=& \half (|\EAco|^2 - |\EAcross|^2 + |\EBco|^2 - |\EBcross|^2)
 \nonumber \\
M_{QU} &=& \half(\EAco\EAcross^*+ \EBco\EBcross^*) + \mathrm{c.c.} \nonumber \\
M_{QV} &=& \half i(\EAco\EAcross^*-\EBco\EBcross^*) + \mathrm{c.c.} \nonumber\\
M_{UT} &=& \half(-\EAco\EBcross^*+\EAcross\EBco^*)+\mathrm{c.c.} \nonumber \\
M_{UQ} &=& \half(-\EAco\EBcross^*-\EAcross\EBco^*)+\mathrm{c.c.} \nonumber \\
M_{UU} &=& \half(\EAco\EBco^* -\EAcross\EBcross^*)+\mathrm{c.c.} \nonumber \\
M_{UV} &=& \half i(\EAco\EBco^*+\EAcross\EBcross^*)+\mathrm{c.c.} \nonumber \\
M_{VT} &=& \half i(\EAco\EBcross^*-\EAcross\EBco^*)+\mathrm{c.c.} \nonumber \\
M_{VQ} &=& \half i(\EAco\EBcross^*+\EAcross\EBco^*)+\mathrm{c.c.} \nonumber \\
M_{VU} &=& \half i(-\EAco\EBco^*+\EAcross\EBcross^*)+\mathrm{c.c.} \nonumber \\
M_{VV} &=& \half(\EAco\EBco^*+\EAcross\EBcross^*)+\mathrm{c.c.} \, .
\label{eq:muller10}
\end{eqnarray}
Expressions for the complex \Muller matrices, obtained from linear
combinations of the components above, are given in Appendix~\ref{app:a}.

With suitable normalization, the (complex) Stokes vector for the
signal at the input to the receiver when the beam is translated by
$\vx$ is
\begin{equation}
\vp_{\mathrm{obs}}(\vx) = \int \ud^2 \vx' \, \mM(\vx') \vp(\vx + \vx').
\label{eq:muller11}
\end{equation}
If we first rotate the beam by $\psi$ (from $x$ towards $-y$) and then
translate by $\vx$, the observed signal after back-rotating to the sky basis
is
\begin{equation}
\vp_{\mathrm{obs}}(\vx;\psi) = \int \ud^2 \vx' \, \mLam^\dagger(\psi)
\mM(\mR_\psi^{-1} \vx') \mLam(\psi) \vp(\vx + \vx'),
\label{eq:muller12}
\end{equation}
where $\mR_\psi$ generates a rotation through $\psi$. In general, the
behaviour of the observed polarization under rotations of the instrument
depends not only on the spin of the appropriate \Muller matrix terms
but also on the beam shapes.

For theoretical work it is convenient to decompose the beam \Muller fields
into components that transform irreducibly under rotation of the instrument.
For co-polar main beams that are approximately Gaussian, the following
expansion is particularly useful:
\begin{eqnarray}
\mM(\vx) &=& \frac{1}{2\pi\sigma^2}
\sum_{mn} \mM_{mn} \sigma^{m+n} (\partial_x + i\partial_y)^m
(\partial_x - i \partial_y)^n \nonumber \\
&& \mbox{} \hspace{0.1\textwidth}\times  e^{-\vx^2/2\sigma^2},
\label{eq:muller13}
\end{eqnarray}
where the sum is over integers $m,\, n \geq 0$, and $\sigma$ is the
nominal beam width. We show in
Appendix~\ref{app:b} that this is related to a Gauss-Laguerre expansion
and detail the inversion to obtain the matrix-valued coefficients $\mM_{mn}$.
Note that the symmetries in equation~(\ref{eq:muller6}) hold for the
irreducible components if we interchange $m$ and $n$, for
example $[M_{mn}]_{P^* T} =[M_{nm}]_{P T}^*$.
The $m=n=0$ components describe pure Gaussian \Muller fields;
their effect is the same as a receiver \Muller matrix acting after
ideal Gaussian optics.
Under a rotation of the instrument, we have
\begin{eqnarray}
\mLam^\dagger(\psi) \mM(\mR_\psi^{-1} \vx) \mLam(\psi) &=&
\frac{1}{2\pi\sigma^2}
\sum_{mn} [e^{i(m-n)\psi}\sigma^{m+n} \nonumber \\
&\times&
\mLam^\dagger(\psi) \mM_{mn} \mLam(\psi) \nonumber \\
&\times& (\partial_x + i\partial_y)^m (\partial_x - i \partial_y)^n]
e^{-\vx^2/2\sigma^2}. \nonumber \\
&& \label{eq:muller14}
\end{eqnarray}
Inserting the beam expansion into equation~(\ref{eq:muller12}) and integrating
by parts we can express the observed polarization in terms of a local
expansion in derivatives of the Gaussian-smoothed polarization field,
$\vp(\vx;\sigma)$:
\begin{eqnarray}
\vp_{\mathrm{obs}}(\vx;\psi) &=& \sum_{mn} [e^{i(m-n)\psi}\mLam^\dagger(\psi)
\mM_{mn} \mLam(\psi) (-\sigma)^{m+n} \nonumber \\
&&\mbox{} \times (\partial_x + i\partial_y)^m (\partial_x - i \partial_y)^n]
\vp(\vx;\sigma) .
\label{eq:muller15}
\end{eqnarray}
The derivative terms have spin $m-n$ plus the intrinsic spin of the
polarization field, and the coefficients $\mM_{mn}$ have spin $n-m$
plus the difference of spin between the fields involved, e.g.\ the
\Muller element $[\mM_{mn}]_{PT}$ has spin $n-m+2$. This expansion
generalizes the local expansion introduced by~\citet{Hu03}. For
optical systems where the non-ideal behaviour can be parametrized by
only a few $\mM_{mn}$ matrices, equation~(\ref{eq:muller15}) leads
to a fast way to simulate the signal component of polarization maps
contaminated by systematics for an arbitrary scan strategy (encoded
in the angles $\psi$). We discuss this further in
Section~\ref{sec:powerspectra}.

\subsection{Polarization calibration with beam rotation}
\label{sec:calibration}
Instrument rotation is a powerful way of reducing the impact of imperfections
in the polarimeter. Structure in $\mM(\vx)$ that is not invariant under
rotation produces systematic effects in the maps of observed Stokes parameters
that can be suppressed by a carefully designed scan strategy, involving
multiple visits to sky pixels in a range of orientations. However, there
remain a class of systematics that are not averaged out in this way: those
for which
\begin{equation}
\mLam^\dagger(\psi) \mM(\mR_\psi^{-1} \vx) \mLam(\psi) = \mM(\vx)
\label{eq:muller16}
\end{equation}
for all $\psi$. It is only these contributions to the \Muller matrix
that would survive in the ideal case of a scan strategy where every pixel
is visited in all orientations, and for which the effective \Muller
matrix is
\begin{equation}
\mM_{\mathrm{sym}}(\vx) \equiv \int \frac{\ud \psi}{2\pi}\,
\mLam^\dagger(\psi) \mM(\mR_\psi^{-1} \vx) \mLam(\psi).
\label{eq:muller17}
\end{equation}
If we focus on the element of $\mM$ that couples a spin-$s'$ field into
a spin-$s$ one, equation~(\ref{eq:muller16}) is satisfied by those
irreducible components for which $s-s'+n-m=0$, i.e.\ those which are spin-0.
Alternatively, if we think of expressing the \Muller fields in terms
of polar coordinates $|\vx|,\,\phi$, where $x+iy = |\vx|e^{i\phi}$,
\begin{eqnarray}
[\mM_{\mathrm{sym}}]_{ss'}(\vx) &=& \int \frac{\ud \psi}{2\pi}\,
e^{i(s'-s)\psi} [\mM]_{ss'}(|\vx|,\phi+\psi) \nonumber \\
&=& e^{i(s-s')\phi} \int \frac{\ud \phi}{2\pi}\,
e^{i(s'-s)\phi} [\mM]_{ss'}(|\vx|,\phi) , \nonumber \\
&& \label{eq:muller18}
\end{eqnarray}
which is just the $s-s'$ (angular) Fourier component.

As an example,
consider the conversion of temperature to polarization which is potentially
very troubling for CMB polarimetry. It is the quadrupole part of $M_{PT}$
($\propto e^{-2i\phi}$)
that generates instrument polarization that transforms like a true polarization
under rotation of the instrument, and for this we have, using equation (\ref{eq:muller15})
\begin{eqnarray}
\Delta P_{\mathrm{obs}}(\vx) &=& (\partial_x -i \partial_y)^2 \nonumber \\
&&\mbox{} \times
\sum_m [\mM_{m\,m+2}]_{PT}\sigma^{2(m+1)} \nabla^{2m} T(\vx;\sigma).
\label{eq:muller19}
\end{eqnarray}
If the summation is real, this is only $E$-mode contamination
which can be tolerated at a relatively higher level than $B$-mode
contamination. If there is no cross-polarization,
inspection of equation~(\ref{eq:app3}) of Appendix~\ref{app:a} shows
that $M_{PT}(\vx)$ is real,
and equation~(\ref{eq:app5}) of Appendix~\ref{app:b} shows that
$[\mM_{m\,m+2}]$ will then be real provided that the quadrupolar part
of $M_{PT}$ has its planes of symmetry aligned with the $x$- and $y$-axes.
Under these conditions,
the leakage of temperature to polarization is pure $E$-mode~\citep{Hu03}.

Axisymmetric optical systems have the property that their leakage
described by $M_{PT}$ is a pure quadrupole and so is not suppressed
by rotation~\citep{2004A&A...420..437C}. To see this, we note that
axisymmetry and reflection symmetry demand that the beam fields
generated by a constant field $\vh$ across the beam-defining element
(e.g.\ a horn) are related to $\vh$ by a tensor-valued field that
can only be constructed from $\delta_{ij}$, $\hat{x}_i \hat{x}_j$
and scalar functions of $|\vx|$. This implies that $\vE_A(\vnhat)$
and $\vE_B(\vnhat)$ can be derived from two radial functions,
$\cle_1(|\vx|)$ and $\cle_2(|\vx|)$ as
\begin{eqnarray}
\vE_A &=& \left(\begin{array}{c}
\cle_1 + \half \cle_2 \cos 2\phi \\ \half \cle_2 \sin 2\phi
\end{array}\right) \nonumber \\
\vE_B &=& \left(\begin{array}{c}
- \half \cle_2 \sin 2\phi \\
- \cle_1 + \half \cle_2 \cos 2\phi
\end{array}\right) .
\end{eqnarray}
For an alternative derivation in the CMB context,
see~\citet{2006astro.ph..7312B}.
The $PT$ \Muller matrix element evaluates to
\begin{equation}
M_{PT}(\vx) = \Re (\cle_1 \cle_2^*) e^{-2i\phi} ,
\end{equation}
which is a quadrupole.
Although the leakage cannot be mitigated
by instrument rotation, it is pure electric~\citep{2004A&A...420..437C}
since $[M_{m\, m+2}]_{PT}$ is purely real.
Note also that the coupling is only to anisotropies in $T$ so a uniform
unpolarized brightness would produce no leakage.

\subsection{Optical errors}
\label{sec:optical}

In this section we illustrate the ideas developed above by analyzing
some simple, but common, beam patterns. Taking the ideal radiation
fields as co-polar Gaussians of width $\sigma$, we consider the true
co-polar beams to have pointing errors, and ellipticities aligned
with the instrument axes, i.e.
\begin{eqnarray}
\EAco &=&
\frac{1}{\sqrt{2\pi\sigma^2(1-e_A^2)}}e^{-\frac{1}{4\sigma^2}\left(\frac{(x-b_{A,x})^2}{(1+e_A)^2}+\frac{(y-b_{A,y})^2}{(1-e_A)^2}\right)} \nonumber\\
\EBco &=&
\frac{1}{\sqrt{2\pi\sigma^2(1-e_B^2)}}e^{-\frac{1}{4\sigma^2}\left(\frac{(x-b_{B,x})^2}{(1+e_B)^2}+\frac{(y-b_{B,y})^2}{(1-e_B)^2}\right)}.
\label{eq:optical1}
\end{eqnarray}
These are normalized so their squares integrate to unity. The errors
can be conveniently reparametrized in terms of the average and
differential ellipticity and pointing errors~\citep{Hu03},
\begin{eqnarray}
\vp  &=& \frac{1}{2\sigma}(\vbA + \vbB)\nonumber\\
\vbd &=& \frac{1}{2\sigma}(\vbA - \vbB)\nonumber\\
e_s  &=& \frac{1}{2}(e_A+e_B)\nonumber\\
q    &=& \frac{1}{2}(e_A-e_B).
\label{eq:optical2}
\end{eqnarray}
Cross-polar beam patterns tend to be design-specific and are not
easily generalized. We gave one example in Section~\ref{sec:calibration}
where we discussed axisymmetric systems; here we consider two simple
toy-models instead based on low-order quasi-optical approximations.
The first model has the cross-polar beams as
co-pointing Gaussians with the same width as the ideal co-polar beam:
\begin{eqnarray}
\EAcross &=&
\frac{\nu_A}{\sqrt{2\pi\sigma^2}}e^{-\frac{x^2+y^2}{4\sigma^2}+i\chi_A} \nonumber\\
\EBcross &=&
\frac{\nu_B}{\sqrt{2\pi\sigma^2}}e^{-\frac{x^2+y^2}{4\sigma^2}+i\chi_B},
\label{eq:optical3}
\end{eqnarray}
where the parameters $\nu_A$ and $ \nu_B$ control the
amplitudes, and $\chi_A$ and $\chi_B$ the phases, relative to the
co-polar beams. In our second model
the cross-polar beams have a line of symmetry along one axis:
\begin{eqnarray}
\EAcross &=& \frac{\nu_A y}{\sqrt{2\pi\sigma^4}}
            e^{-\frac{x^2+y^2}{4\sigma^2}+i\chi_A} \nonumber\\
\EBcross &=& \frac{\nu_B y}{\sqrt{2\pi\sigma^4}}
            e^{-\frac{x^2+y^2}{4\sigma^2}+i\chi_B}.
\label{eq:optical4}
\end{eqnarray}
In all cases the fields are normalized so the integrals of their
absolute squares are $\nu_A^2$ and $\nu_B^2$. As with the co-polar
case, it is useful to reparametrize in terms of average and
differential quantities, in this case the real and imaginary parts
of $\nu_A e^{i\chi_A}$ and  $\nu_B e^{i\chi_B}$ -- the average and
differential components of cross polarization in phase and $\pi/2$
out of phase with the main beams:
\begin{eqnarray}
\nu_{a,R} &=& \half(\nu_A \cos{\chi_A} + \nu_B \cos{\chi_B}) \nonumber \\
\nu_{d,R} &=& \half(\nu_A \cos{\chi_A} - \nu_B \cos{\chi_B}) \nonumber \\
\nu_{a,I} &=& \half(\nu_A \sin{\chi_A} + \nu_B \sin{\chi_B}) \nonumber \\
\nu_{d,I} &=& \half(\nu_A \sin{\chi_A} - \nu_B \sin{\chi_B}).
\label{eq:optical5}
\end{eqnarray}

It is cumbersome to proceed exactly in the presence of
the pointing errors, so instead we expand in the small
systematic parameters. For the pointing and ellipticity
parameters, such a low-order expansion is only useful above the beam scale
$\sigma$. Even small pointing errors (relative to $\sigma$) and ellipticities
can produce large effects below the beam scale. This is a clear driver,
quite apart from the secondary scientific benefits, for increasing the
resolution of CMB polarization experiments. The next-generation of
ground-based $B$-mode experiments have planned resolution $\la 10\,
\mathrm{arcmin}$, so a perturbative treatment should be sufficiently
accurate on the scale of a primordial $B$-mode signal.

The \Muller fields follow from equation~(\ref{eq:muller10}) and
their irreducible components, $\mM_{mn}$,
are given in Appendix~\ref{app:b} to first-order in small parameters.
(The second-order terms that are needed for a consistent polarization power
spectrum calculation are also given; see Section~\ref{sec:powerspectra}.)
From these, we can construct the observed Stokes fields when the instrument
is observing at angle $\psi$ from equation~(\ref{eq:muller15}). The result
for the linear polarization for Gaussian cross-polar beams is
\begin{eqnarray}
P_{\mathrm{obs}} &=& [1 - 2 i \nu_{a,R} + \half e^{-i\psi}
\sigma \plusp (\partial_x - i \partial_y) \nonumber \\
&&\mbox{}
+ \half e^{i\psi}
\sigma \minusp (\partial_x + i \partial_y)
+ \half e^{-2i\psi} \sigma^2 e_s (\partial_x - i \partial_y)^2
\nonumber \\
&&\mbox{} + \half e^{2i\psi}\sigma^2 e_s (\partial_x + i \partial_y)^2]
P(\sigma) \nonumber \\
&&\mbox{}
+ [2i e^{2i\psi} \nu_{d,R} + \half e^{i\psi} \sigma \plusbd
(\partial_x - i \partial_y) \nonumber \\
&&\mbox{}
+ \half e^{3i\psi} \sigma \minusbd
(\partial_x + i \partial_y) + \half \sigma^2 q (\partial_x - i \partial_y)^2
\nonumber \\
&&\mbox{}
+ \half e^{4i\psi} \sigma^2 q (\partial_x + i \partial_y)^2]T(\sigma)
\nonumber \\
&&\mbox{} + 2 e^{2i\psi} \nu_{d,I} V(\sigma) ,
\label{eq:optical6a}
\end{eqnarray}
where ${}_{\pm 1} b_d \equiv \vbd \cdot (\hat{\vx}\pm i \hat{\vy})$
are the spin-$\pm 1$ components of $\vbd$, and similarly for
$\vp$.
In the fiducial orientation, $\psi=0$, this reduces to
\begin{eqnarray}
P_{\mathrm{obs}} &=& [1 - 2 i \nu_{a,R}
+ \sigma\vp \cdot \vnabla + \sigma^2 e_s(\partial_x^2-\partial_y^2)]
P(\sigma) \nonumber \\
&&\mbox{} + [2i \nu_{d,R} + \sigma \vbd \cdot \vnabla +
\sigma^2q(\partial_x^2-\partial_y^2)] T(\sigma)
\nonumber \\
&&\mbox{} + 2 \nu_{d,I} V(\sigma),
\label{eq:optical6}
\end{eqnarray}
in agreement with~\citet{Hu03} if we drop the cross-polar terms.
We see that a pointing offset
in both beams, $\vp$, couples to the polarization gradient,
whilst a
differential pointing error, $\vbd$, couples to the temperature
gradient. Average and differential ellipticity errors, $e_s$ and $q$,
couple to local quadrupole-like patterns in
the polarization and temperature
respectively.
As the cross-polar beams have the same shape as the ideal co-polar ones,
the cross-polar errors couple to
the Stokes fields directly, and not through gradients. Hence, these
errors are similar in form to those introduced by the receiver, and
their effect on polarization
can be represented by a
rotation error, $\omega = -\nu_{a,R}$, temperature leakage,
$\gamma_2 = 2 \nu_{d,R}$,  and circular polarization leakage, $w_1 =
2 \nu_{d,I}$.

For our second, parity-odd, model of cross-polar beams, the
$\nu$ terms in equation~(\ref{eq:optical6a}) should be replaced by
\begin{eqnarray}
\Delta P_{\mathrm{obs}} &=& [e^{-i\psi}\sigma \nu_{a,R}(\partial_x -
i\partial_y) - e^{i\psi}\sigma \nu_{a,R}(\partial_x +i\partial_y) ]P(\sigma)
\nonumber \\
&-& [e^{i\psi}\sigma \nu_{d,R}(\partial_x - i \partial_y)
- e^{3i\psi}\sigma \nu_{d,R}(\partial_x + i \partial_y)] T(\sigma) \nonumber \\
&+& i[ e^{i\psi}\sigma \nu_{d,I} (\partial_x - i \partial_y)
  - e^{3i\psi}\sigma \nu_{d,I} (\partial_x + i \partial_y)] V(\sigma) ,
\nonumber \\
&& \label{eq:optical8a}
\end{eqnarray}
so that, for $\psi=0$, the odd-parity cross-polar beams contribute errors
\begin{eqnarray}
\Delta P_{\mathrm{obs}} &=& - 2 i \nu_{a,R} \sigma \partial_y P (\sigma)
+ 2 i \sigma \nu_{d,R} \partial_y T(\sigma) \nonumber \\
&&\mbox{} + 2 \sigma \nu_{d,I} \partial_y V(\sigma).\label{eq:optical8}
\end{eqnarray}
That is, the average and differential components of cross polarization in
phase with the main beams couple to the gradient of linear
polarization and total intensity in the direction parallel to the
line of symmetry, respectively, and the differential component
$\pi/2$ out of
phase with the main beams couples to the circular polarization in a
similar manner.

For an ideal scan, all terms with $\psi$ dependence in
equations~(\ref{eq:optical6a}) and~(\ref{eq:optical8a})
average to zero. The only terms to remain are then
\begin{equation}
P_{\mathrm{obs}} = P(\sigma) - 2 i \nu_{a,R} P (\sigma) +
\half q \sigma^2 (\partial_x - i \partial_y)^2 T(\sigma) ,
\label{eq:optical10b}
\end{equation}
where the second term on the right is only relevant for the Gaussian
cross-polar case.

\section{Power spectrum analysis for time-invariant systematics}
\label{sec:powerspectra}

In the previous sections we have described the impact of various
systematic errors on the observed Stokes parameters. To assess fully
the cosmological consequences of these errors, we should propagate
them through to power spectra and, indeed, to cosmological
parameters. In this section we consider the impact of systematics on
the $B$-mode power spectrum. Characterizing this spectrum is a major
goal of future CMB polarization experiments and controlling
systematic effects will be of critical importance. The case of
systematic effects whose projection onto Stokes maps can be
described as a statistically isotropic and homogeneous random
process was considered in~\citet{Hu03}. This is useful for
benchmarking and is probably a reasonable approximation for some
errors such as the pointing jitter/nutation felt by many space and
balloon-borne experiments. However, the statistical description is
unrealistic for many other systematic effects. For example, optical
imperfections can reasonably be expected to be time-invariant and
their projection onto the map is systematic and determined only by
the scan strategy. Here, we shall concentrate on such time-invariant
systematics, and develop semi-analytic methods for predicting their
effect on the power spectra for arbitrary scan strategies. This is
useful for assessing the impact of uncalibrated constant errors and
to inform methods for removing the effects of calibrated errors to a
reasonable level. We also consider the raster scan (where the
instrument is always in it fiducial orientation) and an ideal scan,
as defined earlier, as special cases for which simple analytic
results can be found.

The raster and ideal scan treat each pixel identically,
and we
can define effective \Muller matrix fields that are
independent of the position on the sky being observed.
For the raster scan these are just the \Muller fields in the fiducial
orientation; for the ideal scan they are $\mM_{\mathrm{sym}}$.
The observed Stokes maps are simply the convolution of
the \Muller fields with the Stokes fields on the sky and so in Fourier
space:
\begin{equation}
\vp_{\mathrm{obs}}(\vl) = 2\pi\mM_{\mathrm{eff}}(-\vl)\vp(\vl).
\label{eq:ps1}
\end{equation}
Here, $\mM(\vl)$ is the Fourier transform of the \Muller matrix-valued
field. In terms of the irreducible components, $\mM_{mn}$, we have
\begin{equation}
2\pi \mM(\vl) = \sum_{mn} (il\sigma)^{m+n} \mM_{mn} e^{i(m-n)\phi_\vl}
e^{-l^2 \sigma^2 / 2} ,
\label{eq:ps2a}
\end{equation}
and for an ideal instrument $2\pi \mM(\vl) = e^{-l^2 \sigma^2 /2} \mI$
where $\mI$ is the identity matrix.  For these simple scans the
observed fields in the flat-sky limit are realizations of a
statistically-homogeneous but generally anisotropic process, where the
anisotropies arise from the contamination fields. Equation~(\ref{eq:ps1})
allows us to calculate the Fourier transform of the $B$-mode map,
$B(\vl)$, since
\begin{equation}
B(\bmath{l}) = \frac{1}{2i}\left(
   e^{2i\phi_\vl}P(\vl) - e^{-2i\phi_\vl}P^*(\vl)\right)
\label{eq:ps2}
\end{equation}
from equation (\ref{eq:not1}).

To estimate the power spectrum we use a simple
pseudo-$C_l$ estimator approach, giving an estimate of the smoothed
(band-)power spectrum
\begin{equation}
\hat{C}^B_{b,\mathrm{obs}} = \frac{1}{2f_{\mathrm{sky}} \int_b l \, \ud l}
{\int_b \ud^2 \vl \, |B_{\mathrm{obs}}(\vl)|^2}
\label{eq:ps3}
\end{equation}
where $f_{\mathrm{sky}}$ is the fraction of the sky that has been
observed and $b$ labels the band.
We denote the denominator, $ 2f_{\mathrm{sky}} \int_b l \, \ud l$
by $N_b$ since it is a measure of the number of independent
Fourier modes in the band given the sky coverage. The estimator ignores
several important effects, such as leakage from $E$- to $B$-modes as a result
of incomplete sky coverage~\citep*{2002PhRvD..65b3505L}, but here we can
ignore these as their interaction with systematic effects will be of
secondary importance. (In our simulations below we adopt periodic
boundary conditions to avoid the complications of $E$-$B$ mixing.)
We also assume implicitly that any noise bias is removed from
the estimated spectrum, and that the effects of smoothing with
an ideal, symmetric Gaussian beam are taken account of by
multiplying by $e^{l^2 \sigma^2}$.

By averaging over CMB realizations (denoted by angled brackets) we
find the mean recovered $B$-mode spectrum, $\langle \CBobs \rangle$,
as a function of the true band-power spectra, $\CT$, $\CE$, $\CB$ and
$\CTE$ (all other spectra are taken to be zero as the CMB is not
expected to be circularly polarized, and we assume parity is not
violated).
Making use of equations~(\ref{eq:ps1}) and~(\ref{eq:ps2}),
and ignoring the variation of $\mM_{\mathrm{eff}}(\vl)$ over the radial
extent of the band, we find
\begin{eqnarray}
\langle \CBobs \rangle &=& \frac{\pi}{2} \sum_{j,j'=2}^{3}
\int \ud\phi_\vl \,  (-1)^{j+j'}
[\mLam^\dagger(\phi_\vl) \mM_{\mathrm{eff}}(-\vl) \mLam(\phi_\vl) \nonumber\\
&&\mbox{} \times \mF_l \mLam^\dagger(\phi_\vl) \mM_{\mathrm{eff}}^\dagger(-\vl)
\mLam(\phi_\vl) ]_{jj'} ,
\label{eq:ps4a}
\end{eqnarray}
where the sum is over the $P$ and $P^\ast$ elements. We have introduced
the matrix of true power spectra
\begin{equation}
\mF_l = \left( \begin{array}{ccc}
\CT   & -\CTE    & -\CTE   \\
-\CTE & \CE+\CB & \CE-\CB \\
-\CTE & \CE-\CB & \CE+\CB \\
\end{array} \right),
\end{equation}
and have dropped the $V$ Stokes parameter from Stokes vectors and
\Muller matrices.

For a raster scan, the receiver errors
introduced in equation~(\ref{eq:muller7}) give a recovered power
spectrum
\begin{eqnarray}
\langle   \CBobs \rangle &=&
[(1+a)^2+\half(f_1^2 + f_2^2)]\CB \nonumber \\
&+& [4\omega^2 + \half(f_1^2 + f_2^2)]\CE + \half(\gamma_1^2+\gamma_2^2)\CT.
\label{eq:ps4}
\end{eqnarray}
Due to the zeroth-order $\CB$ term, the perturbation is first order in
$a$, but second order in the remaining parameters. Therefore, we must
be careful with any physical systematics that contribute to $a$ only at
second order but to $\omega$, $f$ or $\gamma$ at first order, as their
resulting power spectrum contributions will be of the same order.
For the errors due to the optics we proceed from equations~(\ref{eq:ps2a})
and~(\ref{eq:ps4a}) and the M\"{u}ller-matrix components given in
Appendix~\ref{app:b} (or directly from the perturbations in the
observed fields given in Section~\ref{sec:optical}). To get a consistent
result to second order in the systematic parameters, it is necessary
to retain second-order terms in the isotropic part of $\mM_{PP}$ and
the hexadecapole part of $\mM_{PP^*}$. Although these are sub-dominant effects
in the map domain, they produce second-order corrections to $B(\vl)$ that
are proportional to the true $B$-modes, and so produce second-order effects
in the observed power proportional to $C^B_l$. For Gaussian
cross-polar beams the result is
\begin{eqnarray}
\langle   \CBobs \rangle &=&
[1 - e_s^2(\lssq - \sigma^4 l^4)-q^2(2-\lssq - \lsfth/8) \nonumber \\
&&\mbox{} - \vb_d^2(2+ \lssq)/4 - 2 (\nu_{a,R}^2 + \nu_{a,I}^2)]
\CB \nonumber \\
&&\mbox{}
+ (\vb_d^2 \lssq/4 + q^2 \lsfth/8 + 2\nu_{d,R}^2) \CT \nonumber \\
&&\mbox{} + 4 \nu_{a,R}^2 \CE .
\label{eq:ps5}
\end{eqnarray}
Note that, as the perturbation fields are derivatives of the Stokes
parameter fields for the co-polar errors,
the power spectrum perturbations couple to the
true power spectra via polynomials in $\sigma l$. There is no
contribution for the co-polar pointing error, $\vp$, which is as
expected as, for a raster scan, this simply leads to a global spatial
translation of the Stokes parameter fields. Note also that there is no
generation of $B$-mode power at leading order from $E$ for the raster scan
if the optics are pure co-polar. Quite generally, with no optical
cross-polarization, the coupling
of linear polarization to linear polarization is diagonal in the instrument
basis and, moreover, is equal for $Q$ and $U$ to first
order in beam perturbations. (This follows from inspection of
the relevant \Muller elements in equation~(\ref{eq:muller10}).) The
first-order coupling is through the average of the
absolute squares of the two co-polar beams and, provided this is parity
symmetric, will not produce $B$-modes from $E$ for a raster scan.
The form of the cross-polar errors in equation~(\ref{eq:ps5})
follows directly from that for the receiver case, equation~(\ref{eq:ps4}),
with $\omega =-\nu_{a,R}$, $\gamma_2 = 2\nu_{d,R}$ and the second-order
map term for $a$ given by $a = -(\nu_{a,R}^2 + \nu_{a,I}^2)$.
In the presence of receiver \emph{and} optical errors, the mean
observed power spectrum is the sum of equations~(\ref{eq:ps4})
and~(\ref{eq:ps5}) (with the zero-order $\CB$ term included only once)
plus cross-terms between the receiver and beam errors; to
second-order the cross-terms contribute
\begin{eqnarray}
\langle   \Delta \CBobs \rangle &=& (2\gamma_2 \nu_{d,R}-2w_1 \nu_{d,I}+
8\omega\nu_{a,R})\CB \nonumber \\
&&\mbox{} + 2 \gamma_2 \nu_{d,R} \CT - 8 \omega \nu_{a,R} \CE \nonumber \\
&&\mbox{} + \sigma^2 l^2 f_1 q \CTE / 2 .
\label{eq:ps5b}
\end{eqnarray}
Note that the cross-terms vanish with the cross-polar beams.
The terms entering with $\CT$ and $\CE$ follow simply from making the
replacements $\gamma_2 \mapsto \gamma_2 + 2 \nu_{d,R}$ and
$\omega \mapsto \omega - \nu_{a,R}$ in equation~(\ref{eq:ps4}). The terms
entering with $\CB$ are from the second-order isotropic ($m=n$) parts of the
$PP$ and $P^\ast P^\ast$ elements of the combined \Muller matrix
for the receiver and optics (i.e.\ the product of their matrices). Finally,
the $\CTE$ term comes from correlating the $B$-modes produced from real
$E$-modes by spin-flip ($f_1$) errors in the receiver with those
from differential ellipticity acting on the temperature.

For our odd-parity toy-model for cross-polar beams, the
$\nu$ terms in equation~(\ref{eq:ps5}) are replaced by
\begin{eqnarray}
\langle   \Delta \CBobs \rangle &=&
-(\nu_{a,R}^2 + \nu_{a,I}^2)(2-\sigma^2 l^2) \CB
+ \sigma^2 l^2 \nu_{d,R}^2 \CT \nonumber \\
&&\mbox{}
+(\sigma^2 l^2 \nu_{a,R} b_{d,x} - 2 \sigma^2 l^2 \nu_{a,R} \nu_{d,R})\CTE
\nonumber \\
&&\mbox{} + 2\sigma^2 l^2 \nu_{a,R}^2 \CE ,
\label{eq:ps6}
\end{eqnarray}
correct to second-order in the systematic parameters. Note the
appearance now of the cross-power $\CTE$ that arises from $E$-$B$
conversion due to $\nu_{a,R}$ and $T$-$B$ conversion from $\vbd$ or
$\nu_{d,R}$. If we also include the receiver errors, the only cross
term to arise now is $\sigma^2 l^2 q f_1 \CTE / 2$.

As we have seen in Section \ref{sec:calibration}, an ideal scan in
which each pixel is visited in every orientation suppresses
systematic errors that are not invariant under instrument rotation.
The effective \Muller matrix for this scan is $\mM_{\mathrm{sym}}$,
as defined in equation (\ref{eq:muller17}). Hence, we can calculate
the recovered $B$-mode spectrum in a similar manner as for the
raster scan, but now only contamination fields with the correct spin
will contribute. For Gaussian cross-polar beams
and receiver errors we find
\begin{eqnarray}
\langle   \CBobs \rangle &=&
[(1+a)^2  + 2 \gamma_2 \nu_{d,R} -
2w_1 \nu_{d,I} + 8\omega\nu_{a,R}               \nonumber \\
&& \mbox{} - e_s^2(\lssq - \sigma^4 l^4/2)
           - q^2(2-\lssq - \lsfth/8)            \nonumber \\
&& \mbox{} - \lssq \vp^2/2- \vb_d^2(2+ \lssq)/4 \nonumber \\
&& \mbox{} - 2 (\nu_{a,R}^2 + \nu_{a,I}^2)] \CB \nonumber \\
&& \mbox{} + 4 (\omega - \nu_{a,R})^2 \CE.
\label{eq:ps7}
\end{eqnarray}
The qualitatively new effect here is the appearance of the pointing
error. A fixed (in the instrument frame) average displacement of the
pointing centre from its assumed position leads to a symmetric beam
distortion on averaging over all orientations. For a small pointing
error the tendency is to increase the effective beam size and so
reduce the power spectrum below the beam scale. For the odd-parity
cross-polar case, the $\nu$-dependent terms in
equation~(\ref{eq:ps7}) are replaced by
\begin{equation}
\langle \Delta \CBobs \rangle =
-(\nu_{a,R}^2 + \nu_{a,I}^2)(2-\lssq) \CB.
\label{eq:ps7a}
\end{equation}
These results suggest that, at least to second order in the
parameters considered,  the effects of total-intensity leakage can
be entirely removed from the recovered $B$-mode spectrum by an
appropriately-designed scan strategy.

However, there are other sources of constraints on the scan. In
particular, for the ground-based experiments  we are immediately
concerned with, the strategy needs to be designed to best target
patches of the sky free from foregrounds, and to aid the removal of
atmospheric emission from the data, which generally involves
constant-elevation scans. This will limit the amount of cross
linking in each pixel that can be achieved (i.e. the spread of
$\psi$ over which we can observe each pixel). Since each pixel is
treated differently for a realistic scan, the map-domain effects
cannot be represented by a simple convolution with an effective
\Muller matrix. This raises the possibility that systematic errors
that, for the raster and ideal scans only produce contamination
proportional to the true $B$-modes,  may cause leakage of the $E$
into $B$ due to the non-trivial geometric pattern of the scan. That
is, certain errors
may lead to more significant biases when considered with a realistic
scan than for either of the extreme cases we have so far considered.
Note also that the observed fields
are neither statistically-homogeneous nor isotropic for general scans.

In order to investigate this, we derive the mean estimated power
spectrum for an arbitrary scan strategy, defined by a list of
values of $\psi$ for each pixel, using the \Muller matrix
decomposition introduced in Section \ref{sec:mullerfields}.
Introducing appropriate notation, the pixel at $\vx$ is visited
$N_{\vx}$ times during the scan, and the angle of the instrument
basis relative to the fiducial basis on the $i$th visit is
$\psi_i(\vx)$, where $i$ runs from $1$ to $N_{\vx}$. Averaging over
the observations for each visit, the observed Stokes vector is
\begin{equation}
\vp_{\mathrm{obs}}(\vx) = \frac{1}{N_{\vx}}\sum_{i=1}^{N_\vx}
\vp_{\mathrm{obs}}[\vx;\psi_i(\vx)].
\label{eq:ps8}
\end{equation}
Re-writing this in component form, and using equation
(\ref{eq:muller15}) we have
\begin{eqnarray}
p_{\mathrm{obs},j}(\vx) &=&  \sum_{nmk} \{
(-\sigma)^{m+n} R_{m-n+s_k-s_j}(\vx) [\mM_{mn}]_{jk} \nonumber \\
&& \times (\partial_x + i \partial_y)^m (\partial_x - i \partial_y)^n \}
p_k(\vx;\sigma).
\label{eq:ps9}
\end{eqnarray}
Here, $s_j$ is the spin associated with the $j$th element of the
complex Stokes vector, and
\begin{equation}
R_s(\vx) = \frac{1}{N_\vx} \sum_{i=1}^{N(\vx)} \exp[i s \psi_i(\vx)].
\label{eq:ps10}
\end{equation}
This leads to a mean estimated power spectrum of
\begin{eqnarray}
\langle \CBobs \rangle &=& \frac{1}{8 f_{\mathrm{sky}}\int_b l\, \ud l}
\int_b \frac{\ud^2 \vl}{2\pi}
\sum_{j,j'=2}^3 \Bigl\{
(-1)^{j+j'} e^{i(s_{j'}-s_j)\phi_\vl}
\nonumber \\
&&\mbox{} \hspace{-0.1\textwidth}
\times \sum_{mnm'n'} \sum_{kk'} (-1)^{m+n} [\mM_{m n}]_{j k}\Bigl[
[\mM_{m^\prime n^\prime}]_{j^\prime k^\prime}^* \nonumber \\
&&\mbox{} \hspace{-0.1\textwidth} \times
\int \frac{\ud^2 \vl'}{2\pi} \Bigl(
R_{m-n+s_k-s_j}(\vl-\vl^\prime)
R_{m^\prime-n^\prime+s_{k^\prime}-s_{j^\prime}}^*(\vl-\vl^\prime)
\nonumber \\
&&\mbox{} \hspace{-0.1\textwidth} \times
(i\sigma l^\prime)^{m+n+m^\prime+n^\prime} e^{i
  (m-n-m^\prime+n^\prime+s_k-s_{k^\prime})\phi_{\vl^\prime}}
[\mF_{l'}]_{kk^\prime} \Bigr)\Bigr]\Bigr\}, \nonumber \\
&&\label{eq:ps11}
\end{eqnarray}
where $R_s(\vl)$ is the Fourier transform of $R_s(\vx)$. For a
raster scan, $R_s(\vl)=2\pi \delta(\vl)$ and
equation~(\ref{eq:ps11}) properly reduces to
equation~(\ref{eq:ps4a}) after summing over $m,n,m',n'$ with
equation~(\ref{eq:ps2a}). If the optical errors are smooth enough
that only a few terms $\mM_{mn}$ in the beam \Muller fields are
significant, equation~(\ref{eq:ps11}) is an efficient way to
calculate the mean observed $B$-mode spectrum from a set of
orientations $\psi_i(\vx)$ avoiding the need for simulations. It is
one of the main results of this paper and we will make further use of
it in Section~\ref{sec:tolerances}.

As well as biasing our power spectrum estimates, systematic errors will
alter their covariance structure. There will generally be an
increase in the random error in the estimates, and, for inhomogeneous
scans, additional correlations in the power spectra above those due to the
survey geometry and any inhomogeneities in the instrument noise properties.
These changes may also impact upon
the usefulness of an experiment.
To assess their extent, we can calculate the covariance of
the observed spectrum over realizations,
\begin{eqnarray}
\mathrm{cov}(\CBobs,\hat{C}^B_{b',\mathrm{obs}})
&=& \frac{2}{N_b^2}
\int_b \int_{b'} \ud^2 \vl \, \ud^2 \vl' \,  |\langle
B_{\mathrm{obs}}(\vl) B_{\mathrm{obs}}(\vl^\prime)\rangle |^2 ,
\nonumber \\
&& \label{eq:ps13}
\end{eqnarray}
where we have ignored the non-Gaussianity
of the lens-induced $B$-modes~(\citealt*{2004PhRvD..70d3002S}; \citealt*{SCR}).
For the case of an ideal instrument,
the off-diagonal terms vanish and
\begin{equation}
\var(\CBobs) = \frac{1}{f_{\mathrm{sky}}(\int_b l \ud \, l)^2}
\int_b  (\CB)^2 l\, \ud  l = \frac{2}{N_b}(\CB)^2,
\label{eq:ps14}
\end{equation}
which is the standard result for cosmic variance. If we also include
the effects of homogeneous white noise with a power spectrum $N_b^B$,
this expression still holds, with $\CB \mapsto \CB + N_b^B$.
For the raster and ideal scans
it is quite straightforward to determine analytically the variance
in the presence of the specific systematic errors defined in
previous sections. To avoid over-cluttering our formulae, we ignore
any cross terms between different systematic parameters so the
following expressions are only applicable when all but one of the
parameters are set to zero. These results are given to fourth order
in the parameters but with some necessarily sub-dominant terms
neglected: the co-efficients of each quadratic power spectrum term
are only given to leading order in each parameter.
For a raster
scan, the receiver errors lead to a change in the variance in the
recovered power spectrum of
\begin{eqnarray}
\Delta[\var(\hatC^B_{b,\mathrm{obs}})] &=& \frac{2}{N_b} \{ [4a + 3(f_1^2
    + f_2^2)]\CBsq                                  \nonumber \\
&& \mbox{} + [16\omega^4 + 3(f_1^4 + f_2^4)/8]\CEsq \nonumber \\
&& \mbox{} + 3(\gamma_1^4 + \gamma_2^4)\CTsq /8     \nonumber \\
&& \mbox{} + [8\omega^2 + (f_1^2 + f_2^2)]\CBCE     \nonumber \\
&& \mbox{} + (\gamma_1^2 + \gamma_2^2)\CBCT \},
\label{eq:ps15}
\end{eqnarray}
whilst for the optical errors with Gaussian cross-polar beams, the
change in variance is
\begin{eqnarray}
\Delta[\var(\hatC^B_{b,\mathrm{obs}})] &=& \frac{1}{N_b}\{
    [4e_s^2(2\lsfth - \lssq)                        \nonumber \\
&&\mbox{} +  q^2(\lsfth/2 + 4\lssq -8)              \nonumber \\
&&\mbox{} - (b_{d,x}^2 + b_{d,y}^2)(2 + \lssq)      \nonumber \\
&&\mbox{} - 8(\nu_{a,R}^2 + \nu_{a,I}^2)            \nonumber \\
&&\mbox{} + 6(\nu_{d,R}^4 + \nu_{d,I}^4)] \CBsq     \nonumber \\
&&\mbox{} + 32\nu_{a,R}^4\CEsq                      \nonumber \\
&&\mbox{} + [3q^4 \lseth/64 + 12\nudr^4             \nonumber \\
&&\mbox{} + 7(b_{d,x}^4 + b_{d,y}^4)\lsfth/32]\CTsq \nonumber \\
&&\mbox{} + [q^4(2-4\lssq +5\lsfth/2                \nonumber \\
&&\mbox{} - \lssix/2 + 5\lseth/128)                 \nonumber \\
&&\mbox{} + (b_{d,x}^4 + b_{d,y}^4)
                       (8 -8\lssq + 3\lsfth)/64     \nonumber \\
&&\mbox{} + 16\nuar^2 + 2(\nudi^4+\nudr^4)]\CBCE    \nonumber \\
&&\mbox{} + [q^2\lsfth/2+(b_{d,x}^2+b_{d,y}^2)\lssq \nonumber \\
&&\mbox{} + 8\nu_{d,R}^2]\CBCT                      \nonumber \\
&&\mbox{} + q^3(2\lssq -2\lsfth + \lssix) \CBCET \}.\nonumber \\
&&\label{eq:ps16}
\end{eqnarray}
For the odd-parity cross-polar case, the $\nu$ dependent terms in
equation (\ref{eq:ps16}) are replaced by
\begin{eqnarray}
\Delta[\var(\hatC^B_{b,\mathrm{obs}})] &=& \frac{1}{N_b}\{
    [(\nudr^4 + \nudi^4)(6 -6\lssq + 9\lsfth/4)    \nonumber \\
&&\mbox{} + (\nuar^2 +\nuai^2)(4\lssq-8)]\CBsq     \nonumber \\
&&\mbox{} + 12\nuar^4 \lsfth \CEsq                 \nonumber \\
&&\mbox{} + 11 \nudr^4 \lsfth /2 \CTsq             \nonumber \\
&&\mbox{} + [ (\nudr^4 + \nudi^4)
      (2 - 2\lssq + 3\lsfth/ 4)                    \nonumber \\
&&\mbox{} + 8 \nuar^2 \lssq ] \CBCE                \nonumber \\
&&\mbox{} + 4 \nudr^2 \lssq\CBCT   \}. \label{eq:ps17}
\end{eqnarray}
For an ideal scan, only the receiver error terms in $a$ and
$\omega$ remain, and for the optical errors with Gaussian
cross-polar beams, the change in the variance is given by
\begin{eqnarray}
\Delta[\var(\hatC^B_{b,\mathrm{obs}})] &=& \frac{1}{N_b}\{
    [2e_s^2(\lsfth - 2\lssq)                       \nonumber \\
&&\mbox{}  +  q^2(\lsfth/2 + 4\lssq -8)           \nonumber \\
&&\mbox{} - (b_{d,x}^2 + b_{d,y}^2)(2 + \lssq)  \nonumber \\
&&\mbox{} - 2(p_x^2 + p_y^2)\lssq                  \nonumber \\
&&\mbox{} - 8(\nuar^2 + \nuai^2)] \CBsq   \nonumber \\
&&\mbox{} + 32\nuar^4 \CEsq                   \nonumber \\
&&\mbox{} + 16\nuar^2 \CBCE \}. \label{eq:ps18}
\end{eqnarray}
For the odd-parity cross-polar case, the $\nu$ dependent terms in
equation (\ref{eq:ps18}) are replaced by
\begin{eqnarray}
\Delta[\var(\hatC^B_{b,\mathrm{obs}})] &=& \frac{1}{N_b}
    (\nuar^2 +\nuai^2)(4\lssq - 8) \CBsq.
\label{eq:ps19}
\end{eqnarray}
In Section \ref{sec:tolerances} we assess the relative importance of
the bias and the increase in random error.

We can also investigate the impact of systematic errors on the mean
and variance of the estimated spectra through simulations. For an
arbitrary scan, performing simulations by directly smoothing
simulated maps with the beam in the appropriate orientation for each
observation is computationally intensive. For optical errors that
can be parametrized with only a few low-order irreducible
components, a more efficient method is to evaluate the summation in
equation~(\ref{eq:muller15}) directly with simulated Stokes maps and
their derivatives.\footnote{The derivative maps can be easily formed
in Fourier space.} Alternatively, the simulated maps can be
convolved with the real beams in a small number of orientations,
$\psi$, and these convolved maps can then be interpolated off to
implement the required scan. This second method requires that the
beams do not have rapid angular variations. 
Both methods allow one 
to test the semi-analytic results presented here for the power
spectrum, and, for the variance, allow calculation for arbitrary
scans (which is cumbersome to do analytically).
Having found the observed maps, the power spectrum can be estimated
using equation (\ref{eq:ps3}) and compared to estimates obtained from
the input sky maps. By combining the results for a large number of sky
simulations to estimate the mean and covariance of the recovered
spectra, equivalent 
results to those presented above for the semi-analytic approach are found. 
Importantly, the interpolation method provides a means to check the
validity of low-order expansions in irreducible components.

\section{Biases in cosmological parameters}
\label{sec:biases}
One of the major goals of CMB experiments is to measure or constrain
cosmological parameters. In the present context, the most important parameter
is $r$, and so it is necessary to follow systematic errors through to
$r$. As discussed in Section~\ref{sec:introduction},
it is often necessary to control systematic errors to much better than
the random (instrument noise plus sample variance) errors in the power
spectrum since, typically, many power spectrum estimates are combined into
relatively few cosmological parameters.
It is important to note that the effects we are considering
properly apply to the residual effect of systematics after
any attempt to remove them coherently during data analysis. Dealing with known
receiver errors is reasonably straightforward in the time domain -- for
a recent example application to real data see~\citet{2006astro.ph..6606J} --
but dealing with optical errors is considerably more difficult.

We begin by assuming that no \emph{statistical} correction is made in the power
spectrum for systematic effects. The systematics will then typically
lead to parameter biases that we can estimate as follows. A simple
estimator for $r$, equivalent to the maximum-likelihood estimate for a
Gaussian likelihood, is
\begin{equation}
\hat{r} = \frac{\sum_b\partial_r\CB (\CBobs-\CBlens)/\sigma_b^2}{%
\sum_b(\partial_r\CB)^2/\sigma_b^2},
\label{eq:biases1}
\end{equation}
where the true $\CB$ has been written as a sum from
gravitational waves, $r \partial_r \CB$, and weak gravitational lensing,
$\CBlens$. The variances $\sigma_b^2$ are our best approximation to
$\mathrm{var}(\CBobs)$.
The bias in $r$ is the average shift in position due to systematics,
\begin{equation}
\Delta r_{\mathrm{sys}} = \frac{\sum_b \partial_r \CB \langle \Delta
\CBobs \rangle / \sigma_b^2}{\sum_b (\partial_r \CB)^2/\sigma_b^2} ,
\label{eq:biases2}
\end{equation}
where $\langle \Delta \CBobs \rangle$ is the bias in the power
spectrum, and we can ignore any contribution
from $\Delta \mathrm{var}(\CBobs)$ to $\sigma_b^2$ at leading order.
A large bias in $r$ compared to the ideal random error,
\begin{equation}
\sigma_r =
\left(\sum_b(\partial_r\CB)^2/\sigma_b^2 \right)^{-1/2} ,
\label{eq:biases3}
\end{equation}
would demand that a statistical correction be made in the power
spectrum for the systematics.
In principle, this can be done for
known (but unremoved) systematic effects if the
statistics of the true fields are known a priori, either by simulation
or with the methods developed in this paper. For systematic parameters
whose values are uncertain, it may be possible to fit their effect in the power
spectrum with a few parameters that can be marginalised over when estimating
cosmological parameters. An example of the latter is the
treatment of beam uncertainties in the recent \emph{WMAP}
analysis~\citep{2006astro.ph..3452J}.

However, even if such statistical
corrections can be made accurately, systematic effects will generally still
increase the random errors in parameters. We can estimate this increase
by looking at the variance of the estimator in equation~(\ref{eq:biases1})
in the presence of systematics. Linearising in $\Delta \mathrm{cov}(
\CBobs,\hat{C}^B_{b',\mathrm{obs}})$, we have
\begin{equation}
\Delta \sigma_r \approx \frac{\sum_{bb'} \partial_r \CB
\partial_r C^B_{b'} \Delta  \mathrm{cov}(
\CBobs,\hat{C}^B_{b',\mathrm{obs}})/(\sigma_b^2 \sigma_{b'}^2)}
{2 \left[\sum_b
(\partial_r \CB)^2/ \sigma_b^2\right]^{3/2}} .
\label{eq:biases4}
\end{equation}
The impact of systematic effects will be negligible if
both $\Delta r_{\mathrm{sys}}$ and $\Delta \sigma_r$ are much less
than the `ideal' random error $\sigma_r$.

\section{Applications}
\label{sec:applications}
\subsection{Finding limits}
\label{sec:tolerances}

In this section we set tolerances on the systematic parameters introduced
in Sections~\ref{sec:mullermatrices} and~\ref{sec:optical} such that
the bias and increase in random error on $r$ are below some small fraction
of the ideal random error. We work in the context of next-generation
ground-based CMB polarimeters, and take the fractional threshold to
be $10$ per cent.
This should
ensure that the sensitivity of an experiment to $r$ is not significantly
compromised. The implications of these criteria depend
on the true value of $r$ and also the noise and scan properties of
the instrument. Here we take $r=0.01$, a
realistic limit for the next-generation of CMB polarization experiments.
Of course, if the true $r$ is greater than $0.01$ the impact of systematic
effects will be less.
Given these criteria, it is useful to define,
\begin{equation}
\alpha = \left.\frac{\Delta r_{\mathrm{sys}}}{\sigma_r}\right|_{r=0.01}
\label{eq:biases5}
\end{equation}
and
\begin{equation}
\beta = \left.\frac{\Delta \sigma_r}{\sigma_r} \right|_{r=0.01},
\label{eq:biases6}
\end{equation}
so that we demand $|\alpha| \le 0.1$ and $\beta \le 0.1$.

For the properties of the instrument and survey, we use values
appropriate to upcoming ground-based experiments like
QUIET\footnote{http://quiet.uchicago.edu/} and
Clover\footnote{http://www-astro.physics.ox.ac.uk/research/expcosmology/\\groupclover.html}.
Specifically, we assume around 500 useable detectors each with
$200$-$\mu$Ks$^{1/2}$ NET, and a total integration time of one year
over a survey region $\sim 600\, \mathrm{deg}^2$. For simplicity, we
analyse a single square patch of side $25^\circ$ and adopt periodic
boundary conditions to avoid geometric mixing of $E$- and $B$-modes.
We note that planned ground-based experiments that are not situated at
polar sites will necessarily require a less contiguous survey
geometry than that assumed here, but this will have little effect on
our conclusions. We take the ideal Gaussian beams to have full-width
at half-maximum of $10$ arcmin and use $512^2$ pixels. The noise is
assumed white and the entire array gives a one-year sensitivity of
$3.3$ $\mu\mathrm{K}$-$\mathrm{arcmin}$. We assume that every pair
of detectors suffers exactly the same systematic effect which can be
thought of as the worst-case limit. We adopt a simple cosmological
model consistent with the recent WMAP three-year
analysis~\citep{WMAP3pe}. Under these circumstances, if $r=0.01$
then $\sigma_r = 0.005$ (with no attempt to clean out the
lens-induced $B$-modes) and an ideal instrument might expect a
$2\sigma$ `detection' of such an $r$. We further assume that only
primordial fields are present on the sky. Our simulations include
lens-induced $B$-modes by direct remapping of Gaussian realizations
of the primary CMB polarization with a Gaussian lensing-deflection
field.

\begin{figure}
\begin{center}
  \includegraphics{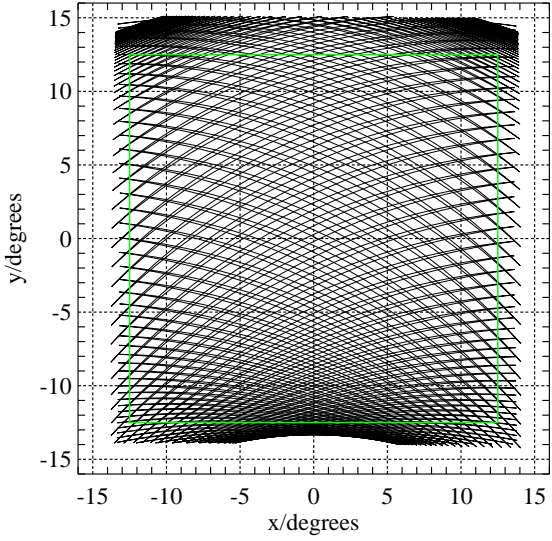}
  \caption[Scan Strategy]{The path of the central focal-plane pixel
  for the semi-realistic scan. The scan is based on a 24-h constant-elevation
  scan from Dome-C. The green (solid) square shows the region used
in the analysis and over which the fields were taken to be periodic.
  \label{fig:scan}}
\end{center}
\end{figure}

We shall consider three different scan strategies: a raster scan, an
ideal scan and a semi-realistic scan. The raster scan is a useful
fiducial scan, and the ideal scan represents the best-case scenario.
The semi-realistic scan allows us to assess the likely impact of
beam rotation on controlling contamination fields. Since the
construction of a realistic scan is very instrument- and
site-specific, here we consider only a toy-model scan that provides
for some variation in the angles $\psi$ between pixels across the
field. The scan path is shown in Fig.~\ref{fig:scan}; it is based on
a constant-elevation scan from Dome-C in Antarctica -- a leading
candidate to site future ground-based CMB experiments.\footnote{The
Clover experiment was originally planned to be sited at Dome-C but
this has since changed to Atacama, Chile. The BRAIN pathfinder
experiment was deployed at DOME-C for the 2005-06 austral summer.}
In using this scan, the effects of differential sky coverage on the
noise are ignored, i.e.\ we assume that each pixel is observed to
the same depth.

\begin{figure}
\centering
\includegraphics{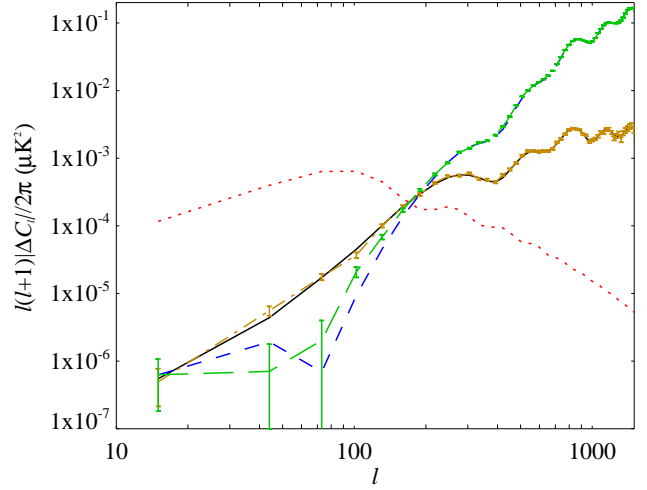}
\caption{Power spectrum perturbations for $b_{d,x}$ and $q$ for the
semi-realistic scan calculated with our semi-analytical method
(black, solid and blue, dashed respectively) and with 1000
simulations (orange, dash-dotted and green, long-dashed respectively). The
error bars are estimates of the theoretical error in the mean of the
1000 simulations due to their finite number. Note the good agreement
between the two methods. For the simulations the scan is
interpolated at intervals of $3.6^\circ$, and it is this
interpolation that leads to the small deviations between the two
methods  evident for $q$ at $l < 200$. The $B$-mode power spectrum
from gravitational waves with $r = 0.01$ is shown in red (dotted) for
comparison.} \label{fig:spectra}
\end{figure}

\begin{figure*}
\centering
\includegraphics{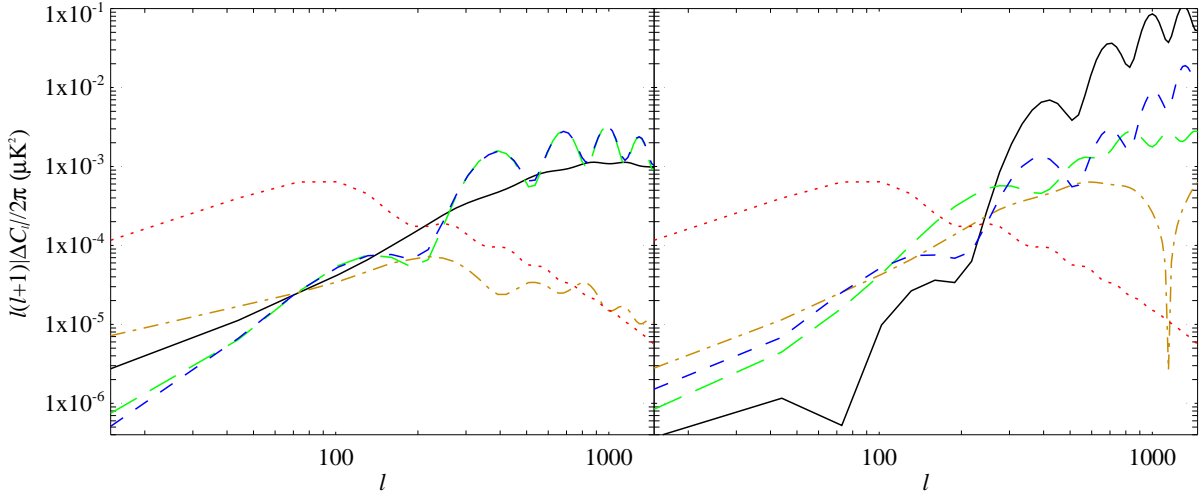}
\caption[Power spectrum for semi-realistic scan]{Power spectra
  perturbations for the semi-realistic scan corresponding to the
  tolerance limits for: (left)  $a$ (black, solid), $\omega$
  (green, long-dashed), $\gamma$ (orange, dash-dotted) and $f$ (blue,
  dashed); and (right) $\nu_{a,R}$ (black, solid), $\nu_{d,R}$ (green,
  long-dashed), $\nu_{a,I}$ (orange, dash-dotted) and $\nu_{d,I}$
  (blue, dashed). The $B$-mode power spectrum from gravitational
  waves with $r = 0.01$ is shown in red (dotted) for comparison.} 
\label{fig:spectra2}
\end{figure*}

\begin{figure*}
\centering
\includegraphics{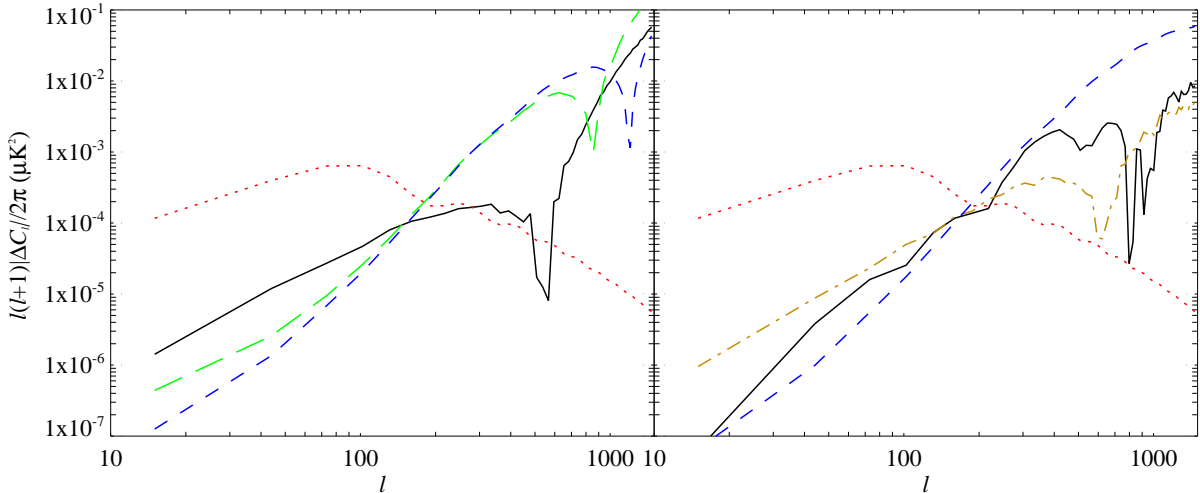}
\caption[Simulated Power Spectrum perturbations]{Simulated power spectra
  perturbations for $e_s$ (left) and $p_x$ (right) for $1000$
  simulations for a raster (green, long-dashed), ideal (blue, dashed) and
  semi-realistic (black, solid) scan. On the left, the perturbation
  for $p_y$ for a semi-realistic scan is also shown (orange,
  dash-dotted). The $B$-mode power spectrum from gravitational waves
  with $r = 0.01$ is shown in red (dotted) for comparison.}
\label{fig:spectra3}
\end{figure*}

We use a combination of the analytic results in
Section~\ref{sec:powerspectra} and simulations to find the mean
change in the power spectrum and its variance and hence find tolerance
limits for each systematic parameter considered in
Sections~\ref{sec:mullermatrices} and~\ref{sec:optical}. We vary
each parameter individually and so do not account for cross-terms
between parameters. The results are summerized in Table \ref{table1}
for our toy-model of odd-parity cross-polar optics. The
corresponding power spectrum perturbations for the semi-realistic scan
are shown in  Figs.~\ref{fig:spectra}--\ref{fig:spectra3}. 
For all parameters apart
from $e_s$ and $\vp$, the power spectrum perturbations and tolerance limits
were evaluated using the semi-analytic and both of the simulation
methods outlined in Section~$\ref{sec:powerspectra}$, with the
different methods showing good agreement. Figure~\ref{fig:spectra}
compares the power spectrum perturbations obtained using the analytic
and interpolation simulation method for a selection of the
parameters for the semi-realistic scan. The good agreement of these
methods validates the 
low-order expansions used in our analytic work, and is representative
of all the parameters considered. Power spectrum perturbations for the
remaining parameters for the semi-realistic scan, calculated with the semi-analytic method, are
shown in Fig.~\ref{fig:spectra2}. For $e_s$ and $\vp$, the assumption
that the parameters are small breaks down below the tolerance limit,
and so we rely on the interpolation simulation method to obtain limits
and power spectrum perturbations (shown in Fig.~\ref{fig:spectra3}) for
these parameters.

\begin{table}
\begin{center}
\caption{Tolerance limits for systematic errors. Three different scan
  strategies are considered: a raster scan, an ideal scan, and a
  semi-realistic scan (see text for details). The
  cross-polar results are for our odd-parity toy-model.}
\label{table1}
\begin{tabular}{ccccc}
\hline
Parameter & Raster & Ideal & Semi-realistic \\
\hline
$a$        &$5.10\times 10^{-3}$&$5.10\times 10^{-3}$&$5.10\times 10^{-3}$\\
$\omega$   &$4.30\times 10^{-3}$&$4.30\times 10^{-3}$&$4.30\times 10^{-3}$\\
$f_1$      &$1.27\times 10^{-2}$&$\infty$            &$5.33\times 10^{-2}$\\
$f_2$      &$1.18\times 10^{-2}$&$\infty$            &$5.23\times 10^{-2}$\\
$\gamma_1$ &$1.61\times 10^{-4}$&$\infty$            &$2.22\times 10^{-4}$\\
$\gamma_2$ &$1.51\times 10^{-4}$&$\infty$            &$2.10\times 10^{-4}$\\
$e_s$      &$5.6 \times 10^{-1}$&$5.5 \times 10^{-1}$&$3.1 \times 10^{-1}$\\
$q$        &$1.20\times 10^{-2}$&$7.26\times 10^{-2}$&$5.39\times 10^{-2}$\\
$p_x$      &$-$                 &$7.3 \times 10^{-1}$&$1.0 \times 10^{-1}$\\
$p_y$      &$-$                 &$7.3 \times 10^{-1}$&$6.0 \times 10^{-1}$\\
$b_{d,x}$  &$2.05\times 10^{-3}$&$1.15\times 10^{-1}$&$2.82\times 10^{-3}$\\
$b_{d,y}$  &$2.05\times 10^{-3}$&$1.15\times 10^{-1}$&$2.84\times 10^{-3}$\\
$\nu_{a,R}$&$2.55\times 10^{-2}$&$7.26\times 10^{-2}$&$2.77\times 10^{-2}$\\
$\nu_{d,R}$&$1.02\times 10^{-3}$&$\infty$            &$1.41\times 10^{-3}$\\
$\nu_{a,I}$&$7.26\times 10^{-2}$&$7.26\times 10^{-2}$&$7.25\times 10^{-2}$\\
$\nu_{d,I}$&$1.12\times 10^{-1}$&$\infty$            &$2.31\times 10^{-1}$\\
\hline
\end{tabular}
\end{center}
\end{table}

In every case, the limiting
criterion is the bias in $r$ which is always at least an order of magnitude
greater than the increase in the random error, $\sigma_r$. This is a
result of the lensing induced $B$-modes dominating the variance in the
power spectrum on large scales. To see this it is useful to write
$\alpha$ and $\beta$ as
\begin{equation}
\alpha \approx \left.\frac{r^2} {2} \sum_b
 \left(\frac{\partial_r \CB}{\sigma_b}\right)^2
  \frac{\langle\Delta \CBobs \rangle}{C_{b,\mathrm{prim}}^B}\right|_{r=0.01}
\label{eq:biases7}
\end{equation}
\begin{equation}
\beta \approx \left.\frac{r^2} {8} \sum_b
 \left(\frac{\partial_r \CB}{\sigma_b}\right)^2
  \frac{\Delta \mathrm{var} (\CBobs)}{\sigma_b^2}\right|_{r=0.01},
\label{eq:biases8}
\end{equation}
where we have made use of equations
(\ref{eq:biases2})--(\ref{eq:biases6}) and the fact that
$\sigma_r|_{r=0.01} \approx r/2$ by design. $C_{b,\mathrm{prim}}^B$
is the primordial $B$-mode power spectrum due to gravitational
waves. Written like this we see that $\alpha$ depends on the ratio
of the bias in the power spectrum to the primordial spectrum, whilst
$\beta$ depends on the ratio of the change in the variance to the
ideal variance in a similar manner. As a result, the large
contribution of gravitational lensing to the $B$-modes suppresses
$\beta$ relative to $\alpha$. Taking an $\omega$ error as an
example, from the results of Section \ref{sec:powerspectra} we find
${\langle\Delta \CBobs \rangle}/{C_{b,\mathrm{prim}}^B} =
4\omega^2\CE/C^B_{b,\mathrm{prim}}$ whilst ${\Delta \mathrm{var}
(\CBobs)}/{\sigma_b^2} = 8\omega^2\CE/\CB$ to leading order in
$\omega$. As $\CB > C_{b,\mathrm{prim}}^B$ at all scales of
interest, $\alpha > \beta$. Note that this assumes we cannot access
the reionization information: for an experiment with sufficient sky
coverage, the reionization peak, which is expected to
dominate the power spectrum at very large scales, will be resolved and
so the increase in random error will be more significant.

As expected, the limits for $a$ and $\omega$ are independent of the
scan strategy, as they simply rescale the $Q\pm iU$ spin states and
the errors they introduce transform like a true polarization. The
limit for $a$ is more strict than might na\"{i}vely be expected, as
the error not only amplifies the primordial spectrum, but also leads
to imperfect subtraction of the lensing contribution. For the raster
scan, there are no limits for an average co-polar pointing error,
$\vp$, as a global displacement has no effect on the power spectrum.
For the ideal scan, $\gamma_1$, $\gamma_2$, $f_1$, $f_2$,
$\nudr$ and $\nudi$ are not constrained since the errors
they produce disappear when averaged over all orientations.

Comparing the semi-realistic scan to the raster scan, many of the
tolerances are relaxed as a result of the cross-linking, but the
improvement is relatively modest. In the case of the tightest
tolerances, those for $\gamma_1$ and $\gamma_2$ which leak
temperature into linear polarization, the limits are only relaxed by
around $40$ per cent, which suggests that our ability to  control
these errors with beam rotation is limited. Although an ideal scan
removes these errors completely, a scan that covers $90$ per cent of
basis orientations only relaxes the limits by a factor of $10$. That
is, the cross-linking needs to be almost complete before we see
significant changes in the tolerances, and so demanding such
improvement would put considerable constraints on the scan strategy.
Such constraints are unlikely to be compatible with other
constraints such as the desire to use constant-elevation scans to
avoid gradients in the average atmospheric signal. Having the
ability to rotate the instrument directly about its boresight,
rather than relying on rotation induced by scanning (and sky
rotation), will likely prove very useful for efficient suppression
of some systematics in future ground-based surveys.

The semi-realistic scan introduces a further new feature over the
raster and ideal scan: not every pixel is seen at the same angles
$\psi$, so there is differential beam rotation across the field.
This gives rise to new effects, such as transformation of $E$-modes
into observed $B$-modes even for an experiment with no
cross-polarization, and identical coupling to the sky and
propagation through the receiver for the two polarizations. Consider
an otherwise ideal instrument with equal elliptical co-polar beams
for the $A$ and $B$ polarizations. If the range of $\psi$ were the
same in every pixel, the observed $Q$ and $U$ maps would simply be
the true maps convolved with some effective beam; this does not
produce any transformation of $E$-modes into $B$. Similar comments
apply to a common pointing error, $\vp$, with otherwise ideal
optics. The effect of differential beam rotation in the errors from
the average ellipticity, $e_s$, and pointing parameters can be seen
in Fig.~\ref{fig:spectra3}, which shows the power spectrum
perturbations for these parameters for all three scan strategies.
For both parameters, transformation of $E$-modes to $B$-modes leads
to an increase in the bias at low $l$. However, this does not lead
to a significant tightening of the tolerances. This is partly
fortuitous, as the $E$-to-$B$ errors from differential beam rotation
contribute with opposite sign to the $B$-to-$B$ errors that the
parameters contribute for an ideal scan. The $E$-to-$B$ effect may
be more significant for a more realistic scan, perhaps requiring
multiple elevation scans to cover the survey region. (Interestingly,
demanding better cross-linking to control temperature leakage, and
suppress low-frequency noise, is likely to result in a more complex
scan strategy with increased differential rotation.) As an extreme
example, we can examine the effects of a scan in which the basis
orientation for each sky pixel is selected at random, as we might
expect this to maximize differential rotation. With such a scan, the
limits become $0.024$ for $e_s$ and $0.058$ for $|\vp|$ which are
still not particularly tight. These results suggests that, provided
we avoid pathologically-constructed scans, transformation of
$E$-modes to $B$-modes from differential rotation is unlikely to be
significant. However, transformation of $E$-to-$B$ through
cross-polarization always warrants careful control.

\subsection{Real Beams}
\label{sec:realbeams}

\begin{figure*}
\begin{center}
  {\includegraphics{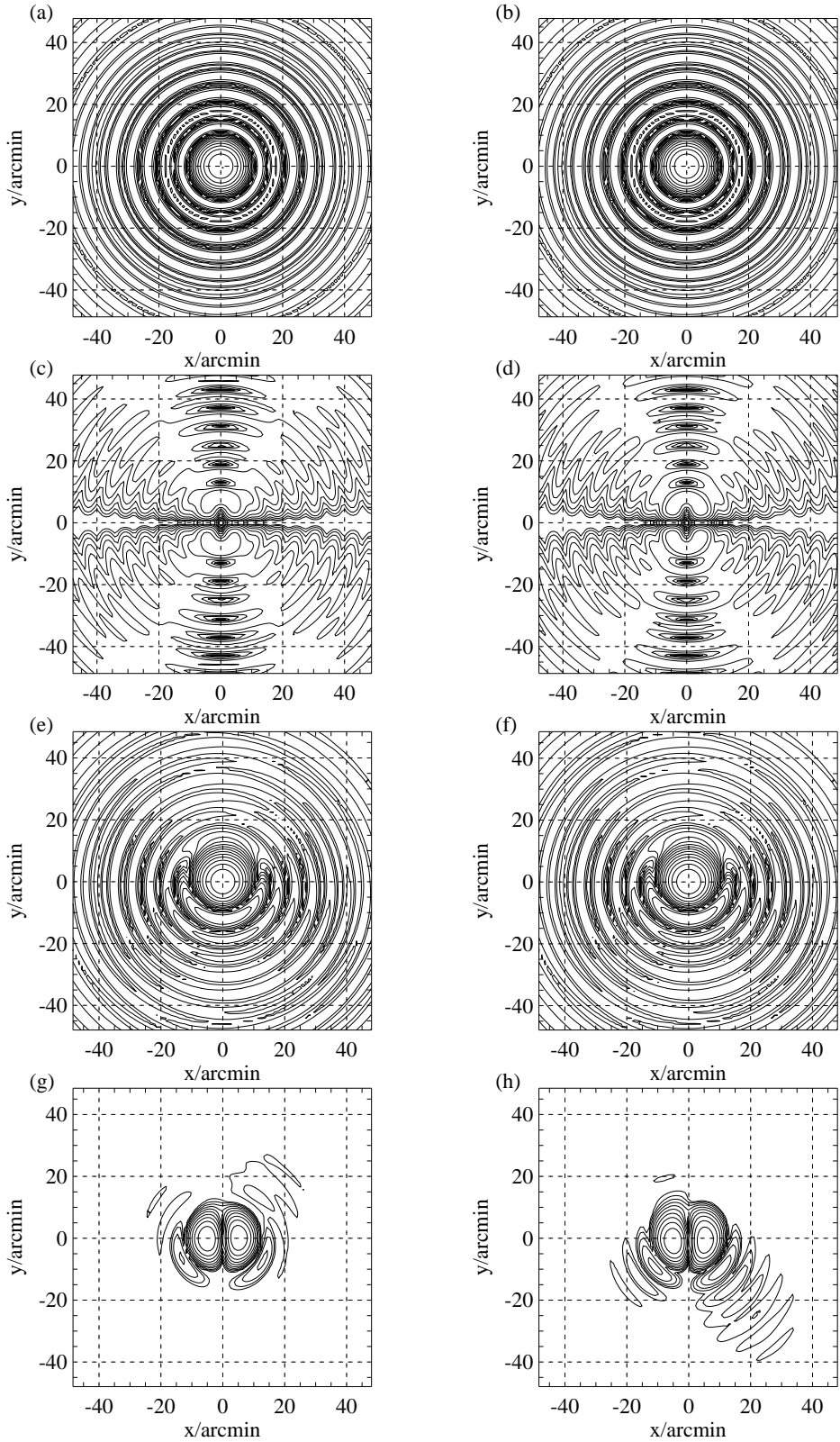}}
  \caption{Simulated (power) beam patterns for a realistic optical
  setup. Panels (a)--(d) show $|\EAco|^2$, $|\EBco|^2$,
  $|\EAcross|^2$ and $|\EBcross|^2$  for a central, on-axis
  pixel, and panels (e)--(h) show the same for a pixel at the edge of
  the focal plane array. The average total power in the cross-polar
  beams, relative to the corisponding co-polar beams is $2.2 \times
  10^{-7}$ for the central pixel, and $2.3 \times
  10^{-6}$ for the shifted pixel.
  The contours are at $-3$ dB intervals, down to $-57$ dB for the
  co-polar beams, and $-30$ dB for the cross-polar beams.}
  \label{fig:graspbeams}
\end{center}
\end{figure*}

So far we have considered the impact of specific beam non-idealities
that can be simply parametrized. However, there may be significant
effects from beam characteristics that are unexpected, or not easily
parametrized and are not present in our simple beam models.
Therefore, in this section we will extend our scope to examine the
effects of real, or simulated, beam patterns. For simplicity, we
will only consider raster and ideal scans. From simulations of the
far-field patterns, $\vE_A$ and $\vE_B$, we can form the beam
\Muller fields using equation~(\ref{eq:muller10}). Then, from
equation~(\ref{eq:ps4a}), we can calculate the bias in the recovered
power spectrum, and from that the bias in $r$.
We simulated beam patterns for a realistic optical setup with a
physical-optics code. (Specifically, the simulations were done in the
context of the Clover experiment.)
Beams
were simulated for three different focal plane positions: the
central pixel and two pixels at the edge of the array, displaced
from the centre along the optical axes. These pixels are expected to
be representative of the entire array. A re-calibration of the beam
centres and axes was performed to remove simulation artifacts. The
power profiles in co- and cross-polarization for the two
polarization states of the central pixel and one edge pixel are
shown in Fig.~\ref{fig:graspbeams} (the remaining edge pixel having
similar features to the one shown). One new feature, not considered
in our previous analysis, that is immediately evident from these
profiles is the subsidiary maxima present in the co-polar beams.
These sidelobe features would not be easy to parametrize, particularly for the
pixels at the focal plane edge where the effect breaks the azimuthal
symmetry of the profiles. When evaluating systematic effects, we
assumed the data analysis would be performed with Gaussian co-polar
beams, with relative powers and beam sizes matched to the simulated
beams. We analysed each pixel separately, but scaled the noise down
to a value appropriate for a 500-element array. This is equivalent
to assuming that every pixel in the focal plane has the same
imperfect optical response. We performed our analysis with an ideal
receiver.

\begin{figure}
\begin{center}
  {\includegraphics{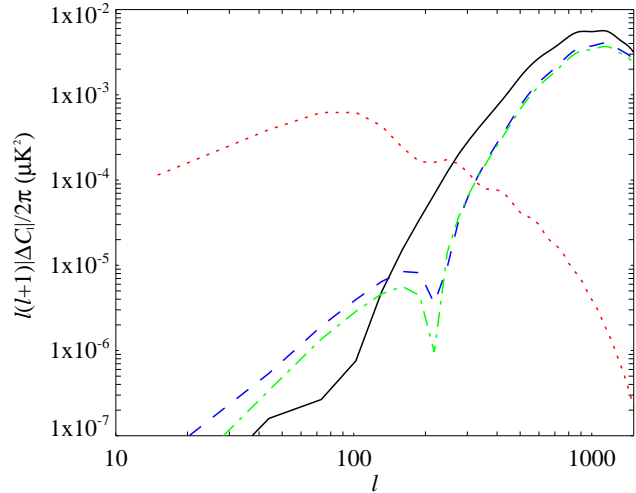}}
  \caption[Power spectrum perturbations for simulated
  beams]{Perturbations to the $B$-mode power spectrum for the
  central focal plane pixel (black, solid) and the edge
  pixels (blue, dotted and green, dash-dotted) for a
  raster scan. (For the edge pixels, the perturbations are negative
  below the kink at $l \sim 200$.) The $B$-mode
  power spectrum from gravitational waves with $r=0.01$ is shown
  in red (dotted) for comparison. }
  \label{fig:graspspectra}
\end{center}
\end{figure}

For the central focal plane pixel $\alpha$ is around $2$ per cent,
irrespective of the scan strategy. Of the pixels considered, this is
the worst case (with $\alpha$ well below the per cent level for the
remaining pixels) and is well within the tolerance limits we have
previously suggested. Furthermore, this bias is mainly a result of
our assumption of simple Gaussian beams in the analysis, and could
be reduced by a more sophisticated treatment of the average,
symmetric co-polar beam profiles. Fig.~\ref{fig:graspspectra} shows
the perturbations to the power spectrum for a raster scan. The
perturbations are largely independent of the scan strategy, and due
almost entirely to the true $B$-modes: the contribution from total
intensity and $E$-modes is always below $10^{-6}~\mu \mathrm{K}^2$.
These results were confirmed by direct convolution of simulated
Stokes maps with the \Muller beams. These simulations also confirmed
that the increase in the random error in $r$ is sub-dominant to the
bias, as expected from Section~\ref{sec:tolerances}.

\section{Discussion}
\label{sec:discussion}

We have presented a framework to describe, in a general manner, the
scientific impact of instrumental systematic errors in CMB
polarimetry experiments surveying small sky areas. A major focus was
the behaviour of systematic errors under instrument rotation. We
introduced spin-weighted \Muller matrices to describe the
propagation of the Stokes parameters through the receiver to
simplify the transformation properties of the systematic effects
under rotations. The optical coupling of the receiver to the
incoming radiation field is described by \Muller matrix-valued
fields. We gave expressions for these in terms of the vector-valued
field pattern of the antenna, and introduced a convenient
decomposition of the matrix fields into components that transform
irreducibly under rotations. The decomposition is useful for
analytical work and also provides an efficient means of simulating
the effect in Stokes maps of systematic effects for an arbitrary
scan strategy.

We compared two popular receiver architectures for CMB polarimetry
and showed that modulating the polarization with a half-wave plate
has the advantage over a pseudo-correlation receiver of
producing no instrumental polarization for ideal optics. It should be
noted that, in the presence of realistic optics, the half-wave-plate
receiver will contribute to instrumental polarization, but at a higher
order in the parameters describing systematics than the
pseudo-correlation receiver. This result 
holds for a waveplate in waveguide -- further analysis is required
for a quasi-optical approach where the waveplate is before the
beam-defining element (and typically is at the beam waist). We also
analysed simple parametrized models for imperfect co- and
cross-polar beams and presented the irreducible components of their
\Muller fields.

We presented an efficient, semi-analytic method for calculating the
bias in the power spectra from time-invariant instrument systematics
described by an arbitrary \Muller response, and for a general scan
strategy. This should prove useful for setting tolerance limits
during the design phase of an experiment. However, since the
analysis is entirely in the map domain, detailed simulations will
still be required for data analysis to account properly for the
interaction of non-trivial map-making with systematic effects. We
also analysed the extreme cases of scans with no and complete beam
rotation and obtained simple results for the biases induced in the
$B$-mode power spectrum, and the increase in the random error in the
power spectrum. Several important effects are missed by these
extreme scans, such as the generation of $B$-modes from $E$-modes,
even in the absence of cross-polarization, by differential beam
rotation across the observed field. It is often important to
propagate the effects of systematic errors through to cosmological
parameters to assess properly their effect on the scientific
returns. We did this for the tensor-to-scalar ratio for a survey
representative of that from next-generation ground-based
instruments, and set tolerance limits for simple parametrized forms
of the receiver and optical systematics. The formalism can also be
applied to non-parametric, simulated or measured beam models and we
illustrated this with the results of physical optics simulations.

Although instrument rotation is potentially a powerful way to
mitigate against stable instrumental systematic effects, we showed
that for a typical scan that can be obtained reasonably from the
ground, the limited range of angles at which a pixel is revisited
limits the extent to which systematics can be controlled. For the
semi-realistic scan considered here we found that tolerance limits
on systematics could be relaxed by between 40 and 400 per cent over
those for a simple raster scan. For all cases considered here, the
tolerance limits for a realistic scan are never significantly
stricter than for a raster scan. The distribution of scan angles
$\psi$ can be improved considerably for constant-elevation scans by
including $z$-axis rotation, rather than relying on sky rotation.
This may prove to be an important element of future campaigns to
detect primary $B$-mode polarization in the CMB.

As previously noted, this paper does not directly consider some
important time-invariant systematic errors. For example, as we have
restricted our analysis to quasi-monochromatic systems, errors such as
bandpass mismatch, and other variations in instrumental response with
frequency have been ignored. However, the methods and analysis
introduced in this paper can be straightforwardly extended to
accommodate such errors by splitting the bandwidth up into a number
of sub-bands in which the instrumental response can reasonably be
modelled as constant, and allowing systematics to vary between the
sub-bands. Each sub-band can be treated independently and the \Muller
matrix used to describe the whole instrument  simply becomes an
appropriately weighted sum over the matrices for each band, and the
analysis can proceed unhindered.

\section{Acknowledgments}

D.O.\ acknowledges a PPARC studentship. A.C.\ is supported by a
Royal Society University Research Fellowship. B.R.J.\ acknowledges
support from a PPARC Postdoctoral Fellowship and an NSF IRFP
Postdoctoral Fellowship. We thank members of the
Clover collaboration for helpful discussions, particularly Paul Grimes
who also provided an early version of the physical-optics simulations
discussed in Section~\ref{sec:realbeams}.

\bibliographystyle{mn2e}
\bibliography{refmnras}

\begin{thebibliography}{}

\bibitem[\protect\citeauthoryear{{Barkats} et~al.}{2005}]{CAPMAP}
{Barkats} D. et~al.,  2005, \apjl, 619, L127

\bibitem[\protect\citeauthoryear{{Barkats} et~al.}{2005}]{2005ApJS..159....1B}
{Barkats} D. et al.,  2005, \apjs, 159, 1

\bibitem[\protect\citeauthoryear{{Bucher}, {Moodley} \& {Turok}}{{Bucher}
  et~al.}{2001}]{iso}
{Bucher} M.,  {Moodley} K.,    {Turok} N.,  2001, Phys. Rev. Lett., 87,
  191301

\bibitem[\protect\citeauthoryear{{Bunn}}{{Bunn}}{2006}]{2006astro.ph..7312B}
{Bunn} E.~F.,  2006, preprint (astro-ph/0607312)

\bibitem[\protect\citeauthoryear{{Carretti}, {Cortiglioni}, {Sbarra} \&
  {Tascone}}{{Carretti} et~al.}{2004}]{2004A&A...420..437C}
{Carretti} E.,  {Cortiglioni} S.,  {Sbarra} C.,    {Tascone} R.,  2004, \aap,
  420, 437

\bibitem[\protect\citeauthoryear{{Efstathiou} \& {Bond}}{{Efstathiou} \&
  {Bond}}{1999}]{1999MNRAS.304...75E}
{Efstathiou} G.,  {Bond} J.~R.,  1999, \mnras, 304, 75

\bibitem[\protect\citeauthoryear{{Feng}, {Li}, {Xia}, {Chen} \& {Zhang}}{{Feng}
  et~al.}{2006}]{2006PhRvL..96v1302F}
{Feng} B.,  {Li} M.,  {Xia} J.-Q.,  {Chen} X.,    {Zhang} X.,  2006, Phys. Rev. Lett., 96, 221302

\bibitem[\protect\citeauthoryear{{Gradshteyn} \& {Ryzhik}}{2000}]{gradshteyn}
{Gradshteyn}, I.~S., {Ryzhik}, I.~M., 2000, Table of integrals, series
and products, 6th edn. Academic Press, New York

\bibitem[\protect\citeauthoryear{{Hu}, {Hedman} \& {Zaldarriaga}}{{Hu}
  et~al.}{2003}]{Hu03}
{Hu} W.,  {Hedman} M.~M.,    {Zaldarriaga} M.,  2003, Phys. Rev. D, 67, 043004

\bibitem[\protect\citeauthoryear{{Hu} \& {Okamoto}}{{Hu} \&
  {Okamoto}}{2002}]{2002ApJ...574..566H}
{Hu} W.,  {Okamoto} T.,  2002, \apj, 574, 566

\bibitem[\protect\citeauthoryear{{Jarosik} et~al.}{2006}]{2006astro.ph..3452J}
{Jarosik} N. et~al., 2006, preprint (astro-ph/0603452)

\bibitem[\protect\citeauthoryear{{Johnson} et~al.}{2007}]{2006astro.ph.11394J}
{Johnson} B.~R. et~al., 2007, \apj, in press (astro-ph/0611394)

\bibitem[\protect\citeauthoryear{{Jones} et~al.}{2006}]{2006astro.ph..6606J}
{Jones} W.~C. et~al., 2006, preprint (astro-ph/0606606)

\bibitem[\protect\citeauthoryear{{Kamionkowski}, {Kosowsky} \&
  {Stebbins}}{{Kamionkowski} et~al.}{1997}]{KKS}
{Kamionkowski} M.,  {Kosowsky} A.,    {Stebbins} A.,  1997, \prd, 55, 7368

\bibitem[\protect\citeauthoryear{{Kovac}, {Leitch}, {Pryke}, {Carlstrom},
  {Halverson} \& {Holzapfel}}{{Kovac} et~al.}{2002}]{DASI}
{Kovac} J.~M.,  {Leitch} E.~M.,  {Pryke} C.,  {Carlstrom} J.~E.,  {Halverson}
  N.~W.,    {Holzapfel} W.~L.,  2002, Nat, 420, 772

\bibitem[\protect\citeauthoryear{{Lewis} \& {Challinor}}{{Lewis} \&
  {Challinor}}{2006}]{2006PhR...429....1L}
{Lewis} A.,  {Challinor} A.,  2006, Phys. Reports, 429, 1

\bibitem[\protect\citeauthoryear{{Lewis}, {Challinor} \& {Turok}}{{Lewis}
  et~al.}{2002}]{2002PhRvD..65b3505L}
{Lewis} A.,  {Challinor} A.,    {Turok} N.,  2002, \prd, 65, 023505

\bibitem[\protect\citeauthoryear{{Lue}, {Wang} \& {Kamionkowski}}{{Lue}
  et~al.}{1999}]{1999PhRvL..83.1506L}
{Lue} A.,  {Wang} L.,    {Kamionkowski} M.,  1999, Phys. Rev. Lett., 83,
  1506

\bibitem[\protect\citeauthoryear{{Masi} et~al.}{2006}]{2005astro.ph..7509M}
{Masi} S. et~al., 2006, \aap, 458, 687

\bibitem[\protect\citeauthoryear{{Montroy} et~al.}{2006}]{boomerangEE}
{Montroy} T.~E. et~al., 2006, \apj, 647, 813

\bibitem[\protect\citeauthoryear{{Page} et~al.}{2006}]{WMAP3pol}
{Page} L. et~al., 2006, preprint (astro-ph/0603450)

\bibitem[\protect\citeauthoryear{{Readhead} et~al.}{2004}]{CBIpol}
{Readhead} A.~C.~S. et~al., 2004, Sci, 306, 836

\bibitem[\protect\citeauthoryear{{Scannapieco} \& {Ferreira}}{{Scannapieco} \&
  {Ferreira}}{1997}]{1997PhRvD..56.7493S}
{Scannapieco} E.~S.,  {Ferreira} P.~G.,  1997, \prd, 56, 7493

\bibitem[\protect\citeauthoryear{{Smith}, {Hu} \& {Kaplinghat}}{{Smith}
  et~al.}{2004}]{2004PhRvD..70d3002S}
{Smith} K.~M.,  {Hu} W.,    {Kaplinghat} M.,  2004, \prd, 70, 043002

\bibitem[\protect\citeauthoryear{{Smith}, {Challinor} \& {Rocha}}{{Smith}
  et~al.}{2006}]{SCR}
{Smith} S.,  {Challinor} A.,    {Rocha} G.,  2006, \prd, 73, 023517

\bibitem[\protect\citeauthoryear{{Spergel} et~al.}{2006}]{WMAP3pe}
{Spergel} D.~N. et~al.,  2006, preprint (astro-ph/0603449)

\bibitem[\protect\citeauthoryear{{Wu} et~al.}{2007}]{2006astro.ph.11392W}
{Wu} J.~H.~P. et~al., 2007, \apj, in press, (astro-ph/0611392)

\bibitem[\protect\citeauthoryear{{Zaldarriaga}}{{Zaldarriaga}}{1997}]{Zalri}
{Zaldarriaga} M.,  1997, \prd, 55, 1822

\bibitem[\protect\citeauthoryear{{Zaldarriaga} \& {Seljak}}{{Zaldarriaga} \&
  {Seljak}}{1997}]{ZS97}
{Zaldarriaga} M.,  {Seljak} U.,  1997, \prd, 55, 1830

\bibitem[\protect\citeauthoryear{{Zaldarriaga} \& {Seljak}}{{Zaldarriaga} \&
  {Seljak}}{1998}]{1998PhRvD..58b3003Z}
{Zaldarriaga} M.,  {Seljak} U.,  1998, \prd, 58, 023003

\bibitem[\protect\citeauthoryear{{Zaldarriaga}, {Spergel} \&
  {Seljak}}{{Zaldarriaga} et~al.}{1997}]{ZSS}
{Zaldarriaga} M.,  {Spergel} D.~N.,    {Seljak} U.,  1997, \apj, 488, 1

\end{thebibliography}

\appendix

\section{Complex beam M\"{U}ller fields}
\label{app:a}

We can extract the complex beam \Muller fields from the real
components in equation~(\ref{eq:muller10}). Those fields for which
the second index has non-zero spin are best expressed in terms of
the spin-$\pm 1$ components of $\vEA$ and $\vEB$:
\begin{eqnarray}
{}_{\pm 1}E_A &\equiv& \vEA  \cdot (\hat{\vx}\pm i\hat{\vy}) =
\EAco \mp i \EAcross \nonumber \\
{}_{\pm 1}E_B &\equiv& \vEB  \cdot (\hat{\vx}\pm i\hat{\vy}) =
-\EBcross \mp i \EBco,
\label{eq:app1}
\end{eqnarray}
whereas those with spin-$0$ component can be expressed directly in
terms of scalar products of $\vEA$, $\vEB$ and their duals
\begin{equation}
\dualvEA \equiv -\left(\begin{array}{c} \EAcross \\ \EAco \end{array}
\right) , \quad
\dualvEB \equiv \left(\begin{array}{c} -\EBco \\ \EBcross \end{array}
\right) , \quad
\label{eq:app2}
\end{equation}
which are obtained by a right-handed rotation through 90\degr\ about
the radiation propagation direction. The independent components of
the complex \Muller\ fields are
\begin{eqnarray}
M_{TT} &=& \half (|\vE_A|^2 + |\vE_B|^2) \nonumber \\
M_{TP} &=& \quart ({}_1 E_A {}_{-1}E_A^* + {}_1 E_B {}_{-1}E_B^* ) \nonumber \\
M_{TV} &=& \half i (\dualvEA \cdot \vEA^* + \dualvEB \cdot \vEB^*) \nonumber \\
M_{PT} &=& \half (\vEA+i\vEB)\cdot (\vEA-i\vEB)^* \nonumber \\
M_{PP} &=& \quart ({}_1 E_A + i {}_1 E_B)({}_{-1}E_A-i{}_{-1}E_B)^* \nonumber\\
M_{PP^*} &=& \quart ({}_{-1} E_A + i {}_{-1} E_B)({}_{1}E_A-i{}_{1}E_B)^* \nonumber\\
M_{PV} &=& \half i(\dualvEA+i\dualvEB)\cdot (\vEA-i\vEB)^* \nonumber \\
M_{VT} &=& \half i (-\vEA\cdot \vEB^* + \vEB \cdot \vEA^*) \nonumber \\
M_{VP} &=& -\quart i ({}_1 E_A {}_{-1}E_B^* - {}_1 E_B {}_{-1}E_A^* ) \nonumber
\\
M_{VV} &=& \half (\dualvEA \cdot \vEB^* - \dualvEB \cdot \vEA^*) .
\label{eq:app3}
\end{eqnarray}
Expressed this way, the transformation properties under a rotation
of the coordinate basis on the sky are manifest.

\section{Beam expansion}
\label{app:b}

The expansion of the beam in equation~(\ref{eq:muller13}) can be
written as
\begin{eqnarray}
\mM(\vx) &=& \frac{e^{-\vx^2/2\sigma^2}}
{2\pi\sigma^2}\sum_{mn} \Bigl[ \mM_{mn} (-1)^n 2^m m! \left(
\frac{x-iy}{\sigma}\right)^{n-m}\nonumber \\
&& \mbox{} \hspace{0.1\textwidth} \times
L_m^{n-m}(\vx^2/2\sigma^2) \bigr],
\label{eq:app4}
\end{eqnarray}
where $L_n^\alpha$ is the Laguerre polynomial~\citep*{gradshteyn},
familiar from the radial wavefunctions of the Hydrogen atom. Each
term in the sum is a Gaussian multiplying a polynomial in $x$ and
$y$. Equation~(\ref{eq:app4}) can be inverted using the
orthogonality of the Laguerre polynomials:
\begin{equation}
\mM_{mn} = \frac{(-1)^n}{2^n n!}\int \ud^2 \vx\,
\mM \left(\frac{x+iy}{\sigma}\right)^{n-m} L_m^{n-m}(\vx^2/2\sigma^2).
\label{eq:app5}
\end{equation}
Note that, for $m>n$ with $m$ and $n\geq 0$,
$L_m^{n-m}(x) = O(x^{m-n})$ so that the integrand in this equation
is always regular at the origin.

In Section~\ref{sec:optical} we introduced a simple parametrization
for offset, elliptical co-polar beams and two toy-model examples of
cross-polar beams. Here we give the $\mM_{mn}$ matrices for these
parametrized beams to first order in the parameters. Generally, if
the maximum order of the polynomials appearing in
equation~(\ref{eq:app4}) is $l$, then there are terms present with
$m$ and $n$ taking all integer values $\geq 0$ such that $m + n \leq
l$. For a first-order expansion of the beam models of
Section~\ref{sec:optical}, $l=2$. Adopting a matrix notation for the
indices $m$ and $n$, we find for the case of Gaussian cross-polar
beams that
\begin{eqnarray}
[\mM_{mn}]_{TT} &=& \left( \begin{array}{ccc}
        1 & - (\plusp /2) & e_s /2 \\
        - (\minusp / 2) & 0 & 0 \\
        e_s/2 & 0 & 0
        \end{array} \right) \nonumber \\
{[\mM_{mn}]}_{TP} &=& \left( \begin{array}{ccc}
        -i \nu_{d,R} & -(\plusbd /4) & q/4 \\
        -(\minusbd /4) & 0 & 0 \\
        q/4 & 0 & 0
        \end{array} \right) \nonumber \\
{[\mM_{mn}]}_{PT} &=&   \left( \begin{array}{ccc}
        2 i \nu_{d,R} & - (\plusbd/2) & q /2 \\
        - (\minusbd/2) & 0 & 0 \\
        q/2 & 0 & 0
        \end{array} \right) \nonumber \\
{[\mM_{mn}]}_{PP} &=&   \left( \begin{array}{ccc}
        1-2 i \nu_{a,R} & - (\plusp/2) & e_s /2 \\
        - (\minusp/2) & 0 & 0 \\
        e_s/2 & 0 & 0
        \end{array} \right) \nonumber \\
{[\mM_{mn}]}_{VV} &=&   \left( \begin{array}{ccc}
        1 & - (\plusp/2) & e_s /2 \\
        - (\minusp/2) & 0 & 0 \\
        e_s/2 & 0 & 0
        \end{array} \right) ; \label{eq:app6}
\end{eqnarray}
the remaining independent, non-zero elements are $[\mM_{00}]_{TV} =
2 \nu_{a,R} = [\mM_{00}]_{VT}$, $[\mM_{00}]_{PV} = 2 \nu_{d,I} = -2
[\mM_{00}]_{VP}$ and $[\mM_{00}]_{VT} = 2 \nu_{a,I}$. Here, for
example, ${}_{\pm 1}p \equiv \vp \cdot (\hat{\vx} \pm \hat{\vy})$
are the spin-$\pm 1$ components of $\vp$. For the power spectrum
analysis in Section~\ref{sec:powerspectra} we need some elements of
$M_{PP}$ and $M_{PP^*}$ to second-order in the systematics as these
appear multiplying the zero-order \Muller matrix. The relevant
second-order, non-zero terms are
\begin{eqnarray}
{[\mM_{00}]}_{PP} &=& -\vb_d^2/4-q^2-(\nu_{a,R}^2+\nu_{a,I}^2) \nonumber \\
{[\mM_{11}]}_{PP} &=& \vp^2/4  + \vb_d^2/8+e_s^2/2-q^2/2 \nonumber \\
{[\mM_{22}]}_{PP} &=& e_s^2/4+q^2/8 \nonumber \\
{[\mM_{04}]}_{PP^\ast} &=& q^2/16 = {[\mM_{40}]}_{P P^\ast} .
\label{eq:app7}
\end{eqnarray}
For the cross-polar beams with a single line of symmetry,
$[\mM_{mn}]_{TT}$ and $[\mM_{mn}]_{VV}$ are unchanged from
equation~(\ref{eq:app6}) but
\begin{eqnarray}
{[\mM_{mn}]}_{TP} &=& -\frac{1}{4}\left( \begin{array}{ccc}
        0 & \plusbd + 2\nu_{d,R} & -q \\
        \minusbd-2\nu_{d,R} & 0 & 0 \\
        -q & 0 & 0
        \end{array} \right) \nonumber \\
{[\mM_{mn}]}_{PT} &=&  -\frac{1}{2} \left( \begin{array}{ccc}
        0 &  \plusbd - 2\nu_{d,R}  & -q  \\
        \minusbd  + 2\nu_{d,R}   & 0 & 0 \\
        -q & 0 & 0
        \end{array} \right) \nonumber \\
{[\mM_{mn}]}_{PP} &=&   -\frac{1}{2}\left( \begin{array}{ccc}
        -2 & \plusp + 2\nu_{a,R}   & -e_s  \\
        \minusp - 2\nu_{a,R}  & 0 & 0 \\
        -e_s & 0 & 0
        \end{array} \right) .  \nonumber \\
&&\label{eq:app8}
\end{eqnarray}
The remaining independent non-zero elements are
$[\mM_{01}]_{TV}=-i\nu_{a,I} = - [\mM_{10}]_{TV}$,
$[\mM_{01}]_{PV}=-i\nu_{d,I} = - [\mM_{10}]_{PV}$,
$[\mM_{01}]_{VT}=-i\nu_{a,I} = - [\mM_{10}]_{VT}$ and
$[\mM_{01}]_{VP}=i\nu_{d,I}/2 = - [\mM_{10}]_{VP}$.
The second-order
terms required for a consistent power spectrum calculation are, in
this case,
\begin{eqnarray}
{[\mM_{00}]}_{PP} &=& -\vb_d^2/4-q^2-(\nu_{a,R}^2+\nu_{a,I}^2) - i p_y
\nu_{a,R}\nonumber \\
{[\mM_{11}]}_{PP} &=& \vp^2/4  + \vb_d^2/8+e_s^2/2-q^2/2\nonumber \\
&&\mbox{} -(\nu_{a,R}^2+\nu_{a,I}^2)/2-i p_y \nu_{a,R}/2\nonumber \\
{[\mM_{22}]}_{PP} &=& e_s^2/4+q^2/8 \nonumber \\
{[\mM_{04}]}_{PP^\ast} &=& q^2/16 = {[\mM_{40}]}_{P P^\ast} .
\label{eq:app9}
\end{eqnarray}

\bsp
\label{lastpage}
\end{document}